\documentclass[aps,twocolumn]{revtex4-1}

\usepackage[utf8]{inputenc}
\usepackage{amsmath}
\usepackage{stmaryrd}
\usepackage{color}
\usepackage{mathrsfs}
\usepackage{yfonts}
\usepackage{rotating}
\usepackage{graphicx}
\usepackage{amsfonts}
\usepackage{mathrsfs}
\usepackage{amsmath}
\usepackage[american]{babel}
\usepackage{wasysym}
\usepackage{slashed}
\usepackage{pgfplots}
\usepackage{braket}
\usepackage{caption}
\usepackage{graphicx}
\usepackage{subfig}
\usepackage{dsfont}
\usepackage{amsthm}
\usepackage{cleveref}
\crefname{proposition}{Proposition}{Propositions}
\crefname{lem}{Lemma}{Lemmas}
\crefname{figure}{Figure}{Figures}
\crefname{corollary}{Corollary}{Corollary}
\crefname{conjecture}{Conjecture}{Conjectures}
\crefname{section}{Section}{Sections}
\crefname{appendix}{Appendix}{Appendixes}
\usepackage{mathtools}

\newcommand{\rrho}[0]{\hat{\rho}}

\newcommand{\ma}[1]{\mathcal{#1}}

\newcommand{\ea}[1]{\begin{align}#1\end{align}}
\newcommand{\eq}[1]{\begin{equation}#1\end{equation}}

\newcommand{\red}[0]{\color[RGB]{221,44,0}}
\newcommand{\green}[0]{\color[RGB]{32,200,83}}
\newcommand{\blue}[0]{\color[RGB]{50,118,210}}
\newcommand{\yellow}[0]{\color[RGB]{254,196,0}}

\DeclareMathOperator{\OA}{OA}
\DeclareMathOperator{\tr}{tr}

\newcommand\oa[1]{\text{OA}\left( #1  \right) }

\newtheorem{proposition}{Proposition}
\newtheorem{theorem}{Theorem}
\newtheorem{corollary}{Corollary}
\newtheorem{conjecture}{Conjecture}

\theoremstyle{definition}
\newtheorem{definition}{Definition}

\usepackage{tikz}
\usetikzlibrary{fit}
\tikzset{%
  highlight/.style={rectangle,rounded corners,fill=red!15,draw,fill opacity=0.5,thick,inner sep=0pt}
}
\newcommand{\tikzmark}[2]{\tikz[overlay,remember picture,baseline=(#1.base)] \node (#1) {#2};}

\tikzset{%
  highlight/.style={rectangle,rounded corners,fill=red!15,draw,fill opacity=0.5,thick,inner sep=0pt}
}

\newcommand{\Highlight}[1][submatrix]{%
    \tikz[overlay,remember picture]{
    \node[highlight,fit=(left.north west) (right.south east)] (#1) {};}
}

\newcommand{\tikzmarkk}[2]{\tikz[overlay,remember picture,baseline=(#1.base)] \node (#1) {#2};}

\tikzset{%
  highlightt/.style={rectangle,rounded corners,fill=blue!15,draw,fill opacity=0.5,thick,inner sep=0pt}
}
\newcommand{\Highlightt}[1][submatrix]{%
    \tikz[overlay,remember picture]{
    \node[highlightt,fit=(up.north west) (down.south east)] (#1) {};}
}

\usetikzlibrary[positioning]
\usetikzlibrary{patterns}
\newcommand{\daywidth}{1.2 cm}

\begin{document}

\title{Cut-resistant links and multipartite entanglement resistant to particle loss}

\author{Gon\c{c}alo M. Quinta}
\email{goncalo.quinta@tecnico.ulisboa.pt}
\affiliation{Instituto de Telecomunicações, Physics of Informations and Quantum Technologies Group, Lisboa, Portugal}
\author{Rui Andr\'{e}}
\email{rui.andre@tecnico.ulisboa.pt}
\affiliation{Centro de Astrof\'{\i}sica e Gravita\c c\~ao  - CENTRA,
Departamento de F\'{\i}sica, Instituto Superior T\'ecnico - IST,
Universidade de Lisboa - UL, Av. Rovisco Pais 1, 1049-001 Lisboa, Portugal}
\author{Adam Burchardt}
\affiliation{Jagiellonian University, Marian Smoluchowski Institute for Physics, \L ojasiewicza 11, 30-348 Krak\'{o}w, Poland}
\author{Karol \.{Z}yczkowski}
\affiliation{Jagiellonian University, Marian Smoluchowski Institute for Physics, \L ojasiewicza 11, 30-348 Krak\'{o}w, Poland}
\affiliation{Center for Theoretical Physics,
Polish Academy of Sciences,
Al. Lotnik\'{o}w 32/46, 02-668 Warszawa, Poland}


\date{August 27, 2019}

\begin{abstract}
In this work, we explore the space of quantum states composed of $N$ particles. To investigate the entanglement resistant to particles loss, we introduce the notion of \textit{$m$-resistant states}. A quantum state is $m$-resistant if it remains entangled after losing an arbitrary subset of m particles, but becomes separable after losing a number of particles larger than m.
We establish an analogy to the problem of designing a topological link consisting of $N$ rings such that, after cutting any $(m + 1)$ of them, the remaining rings become disconnected. We present a constructive solution to this problem, which allows us to exhibit several distinguished $N$-particles states with the desired property of entanglement resistance to a particle loss.
\end{abstract}
\maketitle
\vspace{2mm}

\section{Introduction}

Entanglement is one of the most fundamental resources for quantum technologies. It is the pillar stone of dense coding, quantum teleportation, quantum key distribution
\cite{Ve07}, error correcting codes \cite{LMPZ96}
 and quantum computation \cite{NC00,Jozsa:2003}.
Therefore various aspects of quantum entanglement were investigated \cite{HHHH09}
and  different methods for its experimental detection were proposed \cite{GRW07,GT09}.

After more than two decades of intensive research,
it is fair to say that entanglement in bipartite systems is already well understood and
quantified \cite{HHHH09}.
On the other hand, entanglement in the multipartite set-up is much more
complex and intricate.
 In particular, the topic of characterizing, classifying and quantifying
 the entanglement of many-particle systems
 has proven to be an extremely rich subject
 and  is still an active field of research \cite{BZ17}.
Already for three or more subsystems there is no unique  way to quantify entanglement
 as different entanglement measures induce different orderings and
 are maximized by different states  \cite{Dur:2000,Wu:2001}.
 For instance, the three--qubit state $|\textrm{GHZ}\rangle =(|000\rangle +|111\rangle)/\sqrt{2}$
 is the most entangled with respect to an important three-party
 entanglement measure called three-tangle \cite{CKW00},
 while the state  $|\textrm{W}\rangle =(|100\rangle +|010\rangle +|001\rangle)/\sqrt{3}$
 maximizes the average entanglement contained in two party reductions.

Therefore, it is reasonable to
 characterize entanglement depending on the applications considered.
 A possible example of such a classification comes
 through Stochastic Local Operations assisted with Classical Communications (SLOCC).
 In this approach  \cite{BKMGLS09}, one considers two states to be equivalent if they can be obtained from one another with some finite probability using only local operations assisted with classical communications. States belonging in the same SLOCC class
may then be used to perform the same protocols, which
motivates this very scheme of classification.
 However, although for three qubits all SLOCC classes have been worked out
 \cite{Dur:2000},
 already for four qubits there are infinitely many such classes \cite{Verstraete:2002}.
 It is thus necessary to focus on different traits of quantum states in order to find alternative finite classification schemes.

Many useful and complementary features of quantum entanglement were
 developed in recent years, shedding light into different aspects of multipartite entanglement.
Some approaches  proposed for two subsystems can be directly generalized
for a larger number of parties.
For instance,  the robustness of entanglement,
originally defined for bipartite systems \cite{Vidal:1999} as the
minimal amount of mixing with separable states
needed to wipe out the entanglement, can be directly applied for multipartite systems.

Starting from the notion of the  Bell state one can proceed to the
multipartite domain  by analyzing,
how a given multipartite state is affected if some parts of it are altered or ignored.
Such a procedure leads to the concept of
Absolutely Maximally Entangled (AME) states,
which are maximally entangled with respect to all possible
 bipartitions of the system \cite{Facci:2008,Goyeneche:2015}.
The AME states are particularly relevant for multipartite teleportation and quantum
secret sharing \cite{Helwig:2012}
and for construction of quantum error correction codes \cite{Raissi-2017}.

A natural generalization of AME states are the so called $k$-uniform states,
whereby the partial tracing of any subsystems down to $k$ qudits will result in a maximally mixed state \cite{Scott:2004,Arnaud:2013}.
Such states can be constructed with help of
combinatorial tools like orthogonal arrays \cite{Goyeneche:2014},
which allow for a coarse-grained classification of multipartite
entanglement  \cite{SGZ18}.
Another related measure, called the persistence of entanglement \cite{Briegel:2001},
is defined by the minimal number of local measurements such that
the state becomes completely disentangled
for any measurement outcome.
This property provides a somewhat intuitive notion of entanglement strength,
by looking into the minimal requirements to destroy it completely.

A similar idea was used first by Aravind \cite{Ar97},
who proposed to relate some $N$-partite quantum states
with a link composed of $N$ closed rings.
After any measurement  performed on any subsystem of
the state $|\textrm{GHZ}\rangle$ the remaining bipartite state becomes separable.
Thus this state can be associated with a
particular configuration of three rings,
called {\sl Borromean rings} -- see Fig. \ref{31},
distinguished by the fact that if one ring is cut,
the remaining two become disconnected.  Usage of other tools borrowed from knot theory to analyze multipartite entanglement was further advocated in \cite{KL02,KM19}.
In fact, Borromean rings and their generalizations interacted with various branches of physics, being both an inspiration and a trigger of their development~\cite{Atiyah}.

An alternative interpretation of cutting the rings
was proposed by Sugita \cite{AS07},
who suggested associating individual subsystems  with rings, entangled  states with linked rings,
and the partial trace over a given subsystem with the act of cutting and removing the associated ring.
In such a way, the physical process
corresponding to cutting the ring
does not depend on the outcomes of a measurement
and is defined uniquely \cite{MSS13}.
Such an analogy between quantum entanglement and linked rings,
used later in \cite{QuintaAndre:2018},
will be further explored here.

The aim of this work is to refine  the classification
of multipartite entanglement
by introducing the notion of {\sl $m$-resistant states}.
Entanglement of these $N$-partite states is preserved
even if any $m$ subsystems are traced away,
while removing any further subsystem
makes the state of the remaining $N-m-1$ subsystems
completely separable.
 Despite sharing the spirit of looking into what happens when certain parts of a state are ignored, as for AME states and $k$-uniform states, the above definition does not impose the restriction
of maximal entanglement, which consequently increases the complexity of the problem.

 On the other hand, this approach shares attributes with persistence and robustness of entanglement, in that it measures how the entanglement of a given state resists ignoring a
certain number of non--accessible subsystems,
which corresponds to taking the partial trace over them.
This process is not as invasive as performing
local measurements on selected subsystems
 or mixing the analyzed state with
 locally prepared separable states.

Making use of the analogy between entangled states and
linked rings \cite{AS07}
we investigate first a topological question of
linking $N$ closed rings in such a way that after cutting any $m$ of them the remaining rings are connected, whereas cutting any additional ring separates the remaining 
rings. This problem is solved for any numbers $N$ and $m$
 with use of algebraic techniques developed
in \cite{QuintaAndre:2018} which allow one to
associate to any  link of $N$ rings with a polynomial of $N$ variables.
 A formalism developed for the study of the links can be used not only to gain new intuition and pictorial visualization of the entanglement involved, but also to find representative states for each class of  $m$-resistant quantum states,
 in a way which bypasses a great deal of algebra.

This work is organized as follows. In Section II the topological notion of
$m$-resistant link of $N$ rings is introduced while the analogy to
multipartite quantum entanglement resistant to a loss of a subsystem is
presented in Sec. III.
Exemplary constructions of $m$-resistant states of three and four qubits are
provided in Section IV. In the case of pure states symmetric with respect to
permutation of subsystems it is useful to  use the stellar representation
of Majorana~\cite{AMM10,MGBBB10,RM11}. The search for $m$-resistant
states of $N$ subsystems of local dimension $d$ each
is discussed in Section V,
while the last Section presents
concluding remarks and lists some open questions.
In Appendix we list exemplary quantum states
obtained with use of the combinatorial notion of
orthogonal arrays
and prove a topological proposition on existence
of $m$-resistant links of $N$ rings.

\section{M-resistant links}
\label{MresistLinks}

The question in how many different ways one may link any
 $N$ rings has been addressed in \cite{QuintaAndre:2018}.
 We ignore (or cut) each subset of rings and ask whether the remaining link is connected.
We do not take into account other superfluous details of the link. In this sense, two rings can only be connected in one way because they are either connected or not, while from a knot theory perspective this could be done in infinitely many different ways.

\begin{figure}[h!]
  \centering
    \includegraphics[width=0.30\textwidth]{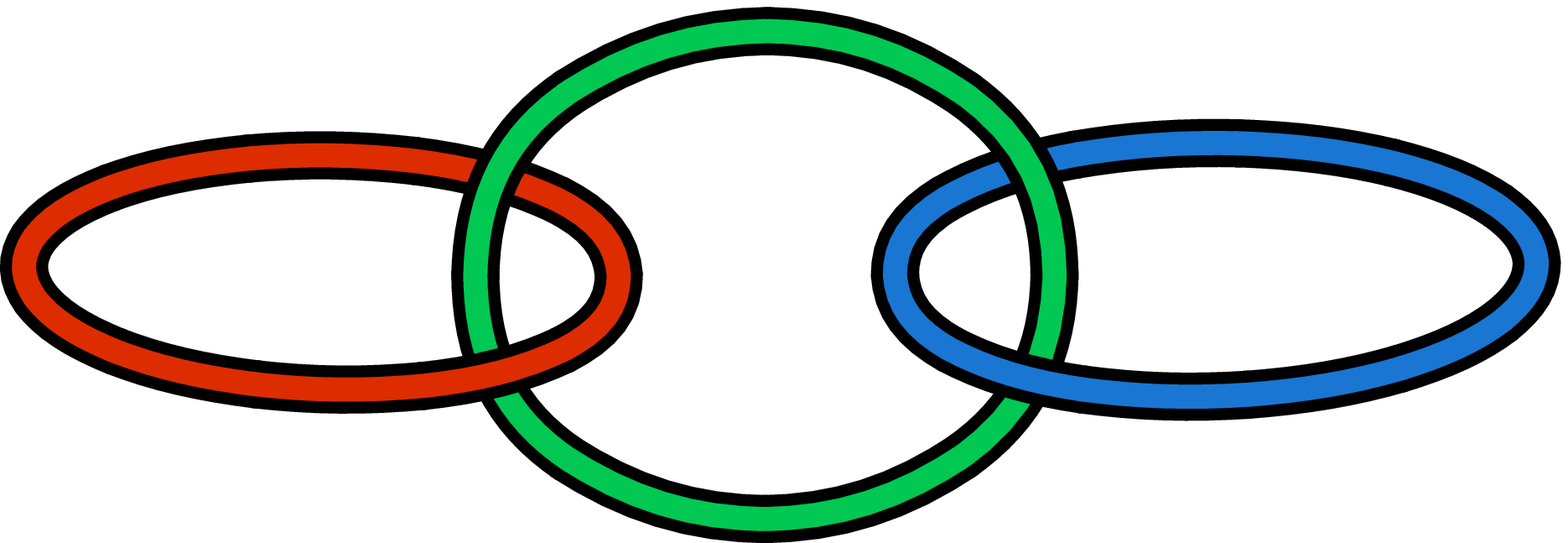}
      \caption{An element of the link class $3^3$, classified by the polynomial ${\red a}{\green b}+{\green b}{\blue c}$, given in (\ref{P33}).}
      \label{33}
\end{figure}

To formalize the above idea, a particular configuration of any number of linked rings is characterized by a polynomial, denoted as $\ma{P}$. The construction of this polynomial starts by associating a variable to each
ring and then interpreting the product between variables to be equivalent to the associated rings being linked. Taking a variable to 0 is then interpreted as cutting or ignoring the associated ring. Thus, a given link can be characterized by the remaining links produced from all possible cuts. As an example, consider the polynomial
\eq{\label{P33}
\ma{P}(3^3) = ab+bc\,.
}
We use the notation $n^i$, introduced in \cite{QuintaAndre:2018}, where $n$ is the number of rings and $i$ is a cataloging index denoting the specific link class. Eq.~(\ref{P33}) implies that the ring $a$ is linked to the
ring $b$, which is also linked with  the ring $c$, while the rings $a$ and $c$ are not directly connected. This can be seen by ignoring the ring $a$, i.e. taking $a=0$, which leaves us with $bc$, meaning the two rings $b$ and $c$ are still linked. The same thing happens when the ring $c$ is ignored, which leaves the term $ab$, implying that the rings $a$ and $b$ remain connected. The link associated with this configuration is represented in Fig.~\ref{33}.

This line of interpretation naturally induces a set of rules for the construction of valid polynomials, which are summarized as follows:
\begin{itemize}
\item[1)] There must not be any repeated terms, i.e. no ring variable can have a power greater than 1 (e.g. $aab$ is the same as $ab$);
\item[2)] Each ring variable must appear at least once;
\item[3)] There must not be first order terms (i.e. we want any ring to be linked with at least another one);
\item[4)] Relabeling of variables is irrelevant;
\item[5)] An $n$-variable monomial $M$ is irrelevant if all of its variables are already present as an $n$-ring link of lesser order monomials, built only with the variables of $M$ (e.g. $abc+ac+ab$ gives the same results as $ac+ab$ after each variable is taken to 0, so we keep only $ac+ab$).
\end{itemize}
These rules for polynomial construction can then be used to count the number of distinct ways to link a given number of rings. For $N=2$, $3$, $4$ and $5$ rings, there are respectively
$1$, $4$, $40$ and $6900$ ways to form a link \cite{QuintaAndre:2018},
while for an arbitrary $N$ the corresponding number has not been found yet.

A particularly interesting subclass of links with $N$ rings is one where all rings remain connected after any $m<N$ cuts, but become fully separated if $m+1$ cuts are performed.
Links with this property will be called {\sl $m$-resistant},
 since they resist up to $m$ cuts before becoming fully disconnected. Clearly, $0\leq m < N-2$, since a link may not resist any cut or it may resist up to $N-2$, in which case we are left with a link of two rings that evidently will not resist any other additional cut.

The problem of finding how many $m$-resistant links of $N$ rings exist is straightforward to solve. There must always be exactly $m+1$ cuts until the link is separated, i.e. until the associated polynomial becomes 0, and it does not matter which rings are chosen to be cut. This immediately implies that the polynomial must be the sum of all possible $N-m$ letter terms. For $N$ rings, there will be $N-1$ classes of different resistances.

\section{Links and quantum states}
\label{MresistStates}

Picking up from the previous section, an analogy can now be made by identifying a ring to a particle and attributing the act of ignoring a ring to the operation of tracing out the particle \cite{AS07}, as if that subsystem has not been detected.
Linked rings are then associated with entangled particles, while separated rings represent separable particles. Distinct links thus represent distinct entanglement classes. Since the trace is used as the connecting ingredient, the entire analogy is basis independent. In addition, since the partial trace is defined for arbitrary Hilbert spaces, we may consider composite quantum systems of any  dimensionality. In this paper, we start working with
$N$-qubit systems,
but in Section V we discuss also
systems with a larger local dimension $d$.

The ``link - quantum entanglement'' analogy goes both ways: one may find the link class associated with a certain state, or; one may look for a state possessing the entanglement properties of a specific link. The former problem is straightforward to answer: one adds a term with $k$ variables to the polynomial whenever the subsystem with the associated $k$ particles is entangled. The latter problem, as we shall show, is much more intricate but can usually be solved using maps between polynomials and quantum states.

The notion of $m$-resistant links of $N$ rings
introduced in the previous section naturally motivates
to distinguish the following class of entangled quantum states
of $N$ subsystems. \\

\textbf{Definition:} ($m$-resistance) An entangled state of $N$ parties is
called $m$-resistant if:
\begin{itemize}
\item It remains entangled as any $m$ of its $N$ subsystems are traced away;
\item It becomes separable if a partial trace is performed over
  an arbitrary set of $m+1$ subsystems.
  \end{itemize}

\begin{figure}[h!]
  \centering
    \includegraphics[width=0.25\textwidth]{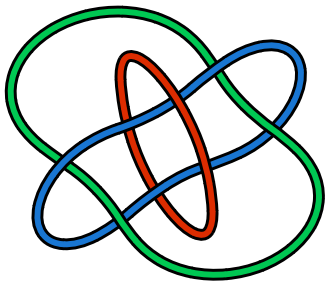}
      \caption{The Borromean link, an example of a 0-resistant link of 3 rings, classified by the polynomial ${\red a}{\green b}{\blue c}$ and given in Eq.~(\ref{P31}).}
      \label{31}
\end{figure}

The first nontrivial example of $m$-resistance arises for $N=3$, which is simple enough to study on intuitive grounds. The case $m=0$ is exemplified by the GHZ state
\cite{Dur:2000,Wu:2001,CKW00}

\eq{
\ket{\rm GHZ} = {1\over \sqrt{2}}\left(\ket{000} + \ket{111}\right) ,
\label{GHZ3}
}
since tracing out any party will result in a separable density matrix. The corresponding link has the properties of the well-know Borromean link, depicted in Fig.~\ref{31}.
This link is constructed in such a way that, if any ring is cut, the remaining two rings become
separated.
 We denote the associated link class by $3^1$, and the polynomial that characterizes it is given by
\eq{\label{P31}
\ma{P}(3^1) = abc\,.
}
For $N$ rings, the generalization is called Brunnian link, with general polynomial
\eq{
\ma{P}(N^1) = abcd\cdots \,,
}
so if a single variable is set to zero, the polynomial vanishes.

\begin{figure}[h!]
  \centering
    \includegraphics[width=0.25\textwidth]{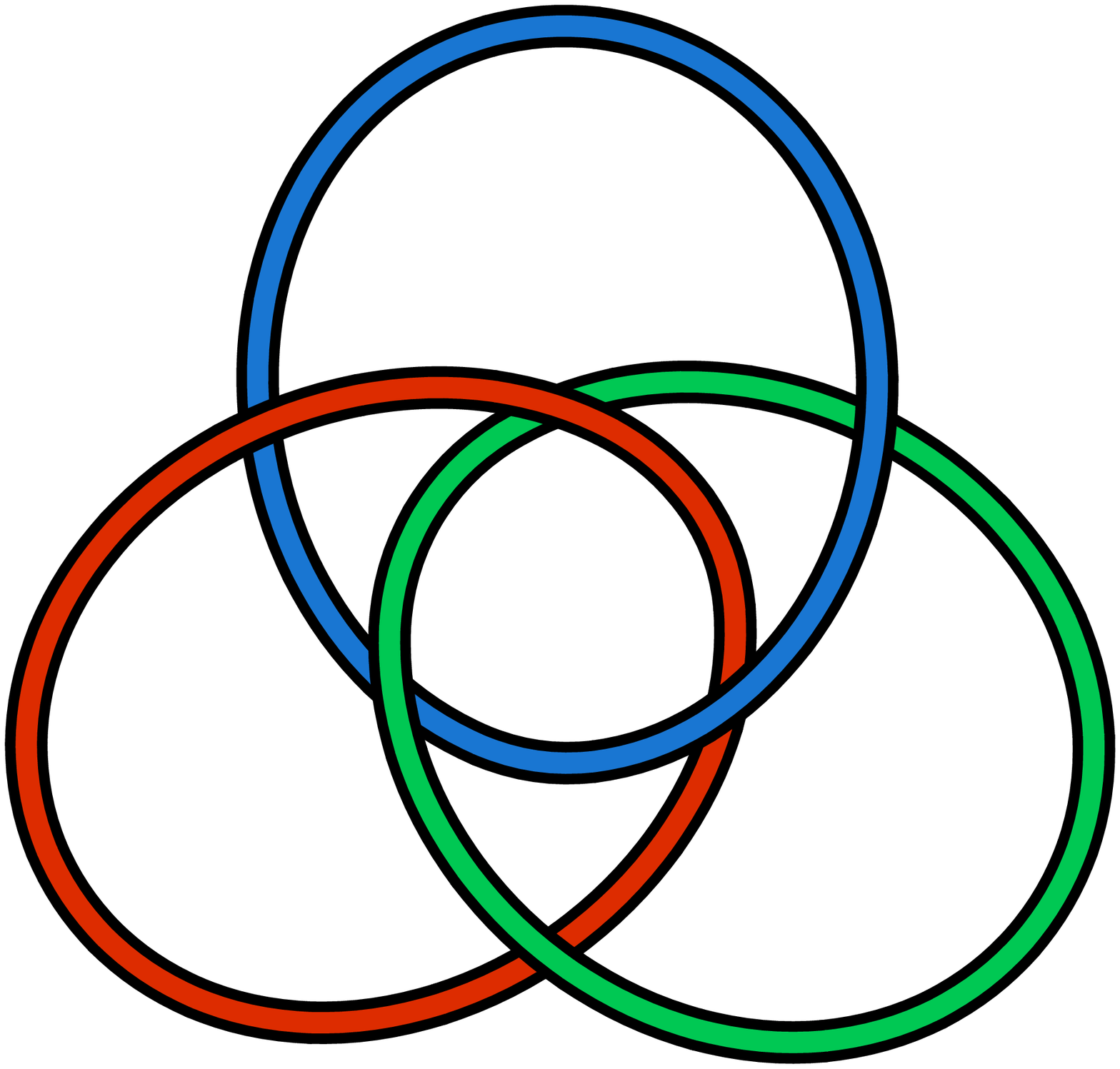}
      \caption{An example of a 1-resistant link of 3 rings, classified by the polynomial ${\red a}{\green b}+{\red a}{\blue c}+{\green b}{\blue c}$, given in Eq.~(\ref{P34}).}
      \label{34}
\end{figure}

The case $m=1$ has different properties, since if any party is ignored
the remaining two parts are still connected.
From the point of view of quantum entanglement
a three-qubit state, called $|\textrm{W}\rangle$  \cite{Dur:2000},
\eq{
\ket{\rm W} = {1\over \sqrt{3}}\left( \ket{100} + \ket{010} + \ket{001} \right)
}
possesses this property,
since partial trace over any of its subsystems will result in an entangled mixed state.
A link which mimics this property is represented in Fig.~\ref{34}.
The polynomial which describes this link is symmetric,
\eq{\label{P34}
\ma{P}(3^4) = ab + ac + bc\,,
}
and the generalization to any number of rings is given by the sum of all two-variable terms.

\section{In search for m-resistant states of N qubits}
\label{search}

The problem of finding $m$-resistant states for a general number $N$ of qubits is not trivial. Since we are only interested in whether a reduced state is entangled or not, for any partial trace, it is clear that at least these states can be symmetric regarding
exchanges of subsystems. This is not strictly necessary, as symmetry under qubit permutation is not invariant under local operations, but it does provide one possible direction. However, the requirement of separability, or lack of it, is not simple to achieve using
any standard algebraic techniques.

In this section we will demonstrate how the analogy between
links of rings and entangled  quantum states
may be used in order to find representative mixed states for all $m$-resistant cases up to $N=7$. For pure states the problem has not been fully solved yet but nevertheless we will present a few examples as well as some geometric intuition behind them.

\subsection{A general rule for mixed qubit states}
\label{GenRule}

The fundamental aspect of the analogy between links of rings and entangled states
is that one can solve certain entanglement problems using the polynomials that characterize the links, and apply 
a map converting polynomials into quantum states.
As we show below, this heuristic reasoning proves to be useful to identify
interesting examples of quantum states resistant to loss of subsystems.

The method to obtain $m$-resistant mixed states of $N$ qubits starts
from the polynomial of $N$ variables
that characterizes the corresponding $m$-resistant links
and consists of  $J$ terms.
For each term $t_i$ of the polynomial, we associate a state $|t_i\rangle\in {\cal H}_2^{\otimes N}\otimes {\cal H}_d$ of the form
\eq{\label{blockform}
|t_i\rangle=
\ket{\phi_{\rm ent}} \ket{\rm \phi_{\rm sep}} \ket{\chi_{\rm D}},
}
where  $\ket{\rm \phi_{\rm sep}}$ stands for a separable state of
$m$ qubits,  while
$\ket{\phi_{\rm ent}}$  denotes an entangled state of
the remaining $N-m$ qubits.
An additional state, $\ket{ \chi_{\rm D}}$
of a single system D with $d$ levels,
plays an auxiliary role, as it will be traced over.
An unnormalized superposition state,
$|\psi_N,m\rangle=\sum_{i=1}^J |t_i\rangle$,
allows us to generate  the resulting mixed state $\hat{\rho}(N,m)$
 of $N$ qubits by partial trace over the $d$-dimensional environment,
\eq{
\rrho(N,m) = {\textrm{tr}_{\rm D} \left[\ket{\psi_{N,m}}\bra{\psi_{N,m}}\right] \over \sqrt{\braket{\psi_{N,m} | \psi_{N,m}}}}\, .
\label{rhoNm}
}
To form an $m$-resistant state we choose
the separable $m$-partite state  $\ket{\rm \phi_{\rm sep}}$  in (\ref{blockform})
to be $\ket{0\ldots 0}$, the extra qudit will assume the index value of the current term, and the entangled part  $\ket{\phi_{\rm ent}}$ will have the form
\eq{\label{Em}
\ket{E_{m}}_{ijk\ldots} = (m+1)\ket{0\ldots0}_{ijk\ldots} + \ket{1\ldots1}_{ijk\ldots}\,,
}
where $ijk\ldots$ are the variables appearing explicitly in the corresponding term of the polynomial. Note that  for any $m>0$ the above superposition is
not symmetric, while for $m=0$ the state
$\ket{E_{0}}_{ijk}$ denotes just the standard $|\textrm{GHZ}\rangle$
 state  (\ref{GHZ3})
among subsystems labeled by indices $i,j$ and $k$.

Putting the above recipe into practice, we have obtained $m$-resistant representative mixed states for all $m$ for the cases $N=3$ up to $7$. In order to test for entanglement/separability of the reduced density matrices after each partial
trace, we used a code available in
\cite{SPv2} which looks iteratively for the nearest separable state and its decomposition into pure product states, returning the distance from the target matrix.

\subsubsection{
$N$=3}
\label{N3}

Although representative pure states have already been provided in the previous section for $N=3$, we show here an alternative answer in terms of mixed states. They are of the general form
\eq{
\rrho(3,m) = {\textrm{tr}_{\rm D} \left[\ket{\psi_{3,m}}\bra{\psi_{3,m}}\right] \over \sqrt{\braket{\psi_{3,m} | \psi_{3,m}}}}\,,
}
with $\ket{\psi_{3,m}}$ specified case by case.

For $m=0$, we have a state with the entanglement properties under partial trace represented in the link of Fig.~\ref{31}. Its polynomial given by Eq.~(\ref{P31}) has a single term only, which leads to the product  state
\eq{
\ket{\psi_{3,0}} = \ket{E_{0}}_{abc}\ket{0}_{d}\,.
}
Hence the resulting state
(\ref{rhoNm}) obtained by partial trace
is pure, and it is shown merely for completeness.
 The remaining case $m=1$ is depicted in Fig.~\ref{34} and is characterized by the polynomial of Eq.~(\ref{P34}). As such, we are lead to the following
 superposition of three states, entangled with respect to different splittings,
\eq{
\ket{\psi_{3,1}} = \ket{E_{1}}_{ab}\ket{0}_c\ket{0}_d + \ket{E_{1}}_{ac}\ket{0}_b\ket{1}_d + \ket{E_{1}}_{bc}\ket{0}_a\ket{2}_d\,.
}

\subsubsection{
$N=4$}
\label{N4}

Using exactly the same procedure of $N=3$, we obtain representative $m$-resistant mixed states for $N=4$, written in general as
\eq{\label{Gen4mixed}
\rrho(4,m) = {\textrm{tr}_{e} \left[\ket{\psi_{4,m}}\bra{\psi_{4,m}}\right] \over \sqrt{\braket{\psi_{4,m} | \psi_{4,m}}}}\,,
}
with $\ket{\psi_{4,m}}$ specified case by case. It is also worth mentioning that the links for more than three rings have been drawn using a technique developed in~\cite{QuintaAndre:2018} that makes direct use of the polynomials associated with them.

\begin{figure}[h!]
  \centering
    \includegraphics[width=0.20\textwidth]{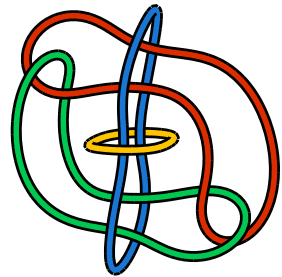}
      \caption{An example of a 0-resistant link of 4 rings, which is a $4$-component Brunnian link, corresponding
    to  the polynomial ${\red a}{\green b}{\blue c}{\yellow d}$ -- see Eq.
     (\ref{PN4k0}).
     Notice that after cutting a single ring all three remaining rings become disconnected.}
      \label{N4k0}
\end{figure}

For $m=0$, the corresponding monomial reads
\eq{\label{PN4k0}
\ma{P}(4^1) = abcd
}
so the state
 associated with the link shown in Fig.~\ref{N4k0}
has the product structure,
\eq{
\ket{\psi_{4,0}} = \ket{E_0}_{abcd}\ket{0}_e\,.
}

\begin{figure}[h!]
  \centering
    \includegraphics[width=0.20\textwidth]{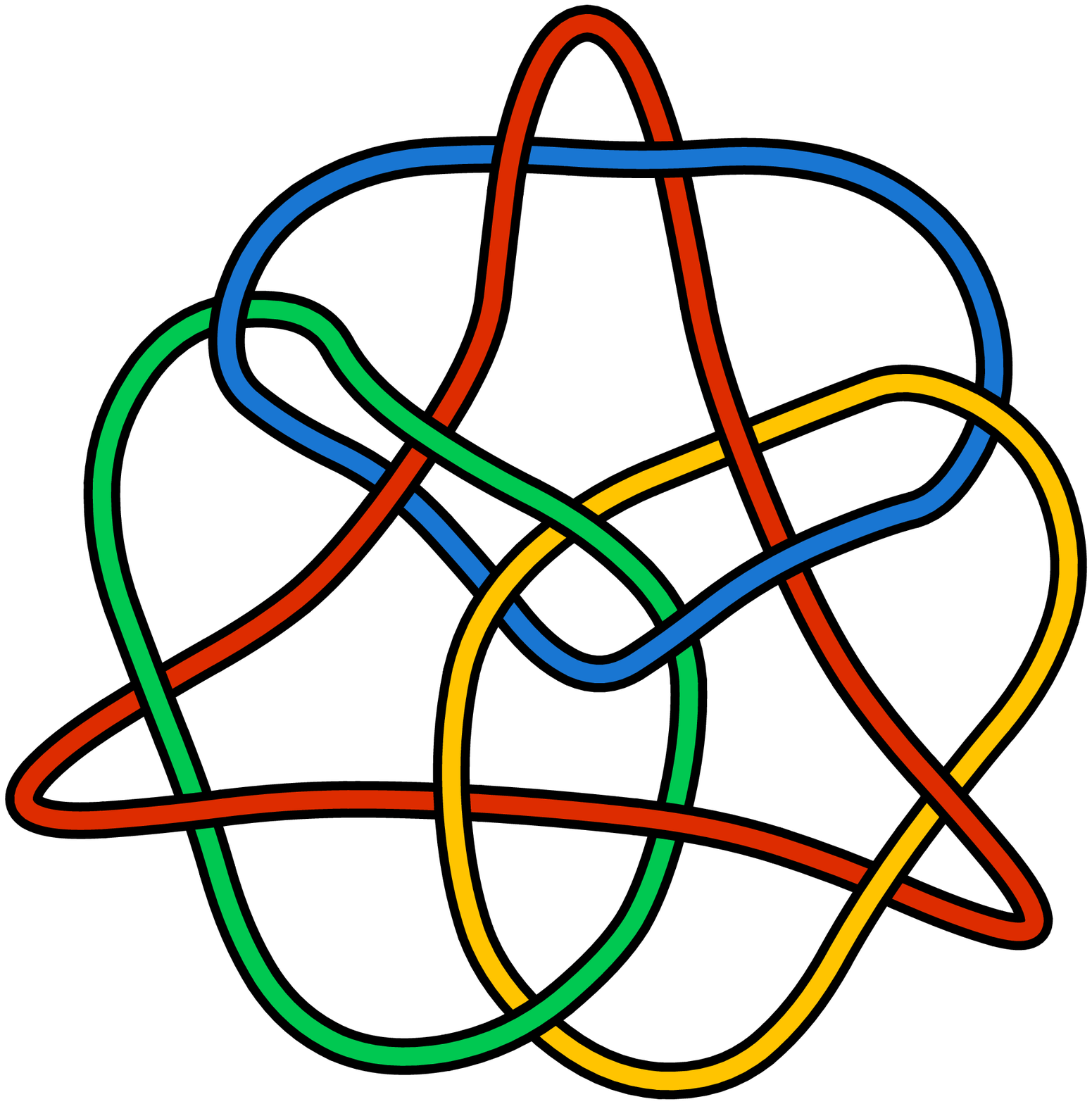}
      \caption{An example of a 1-resistant link of 4 rings, classified by the polynomial ${\red a}{\green b}{\blue c} + {\red a}{\green b}{\yellow d} + {\red a}{\blue c}{\yellow d} + {\green b}{\blue c}{\yellow d}$, given in Eq.~(\ref{PN4k1}).
      After cutting any single ring one arrives at the Borromean link.
      }
      \label{N4k1}
\end{figure}

\begin{figure}[h!]
  \centering
    \includegraphics[width=0.25\textwidth]{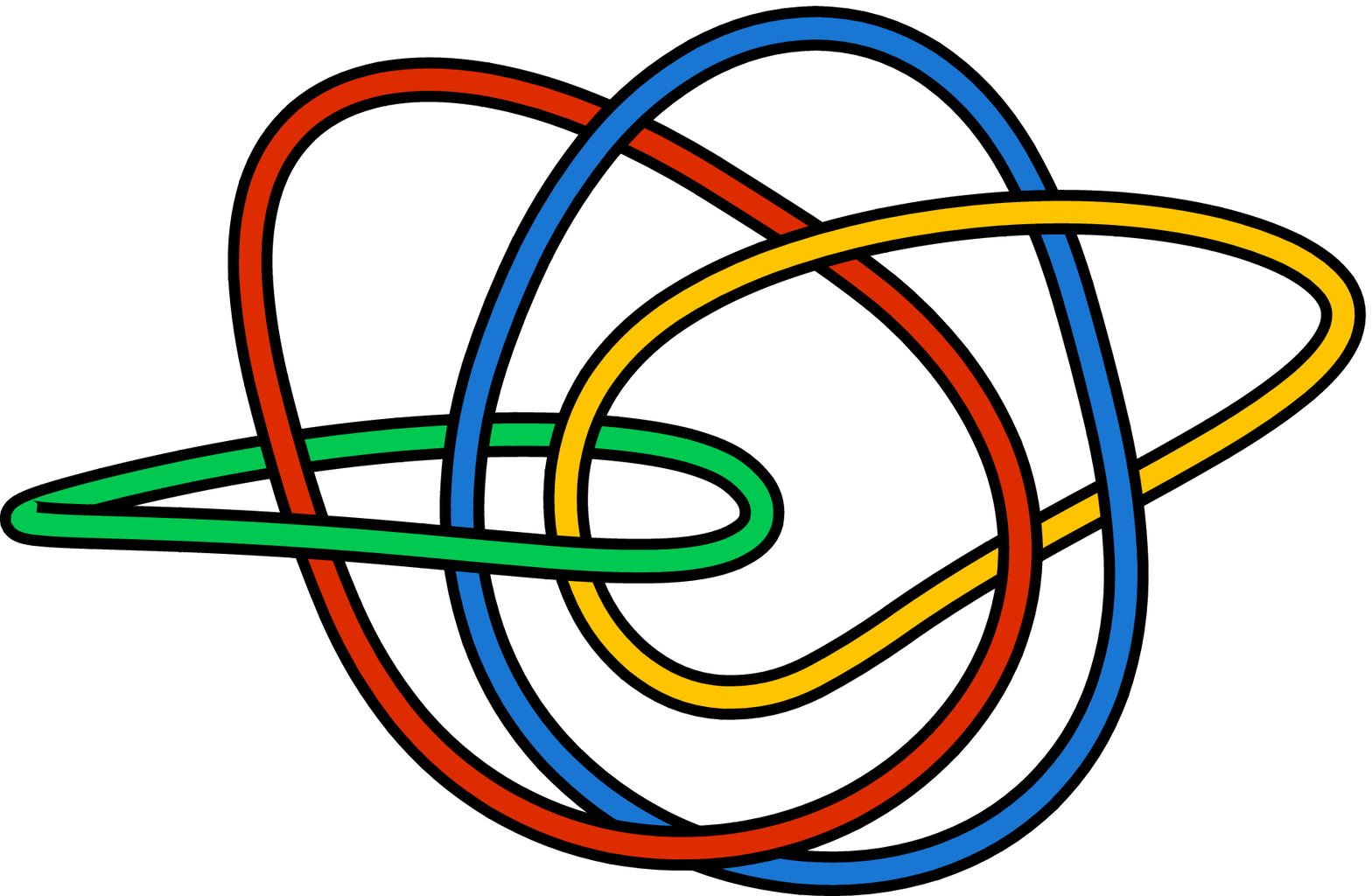}
      \caption{An example of a 2-resistant link of 4 rings, classified by the polynomial ${\red a}{\green b} + {\red a}{\blue c} + {\red a}{\yellow d} + {\green b}{\blue c} + {\green b}{\yellow d} + {\blue c}{\yellow d}$, given in Eq.~(\ref{PN4k2}).
      After cutting any two  rings the remaining two are still connected.
      }
      \label{N4k2}
\end{figure}

The case $m=1$ is associated with the polynomial
\eq{\label{PN4k1}
\ma{P} = abc + abd + acd + bcd
}
which leads to the link of Fig.~\ref{N4k1}.
The polynomial (\ref{PN4k1}) is then mapped into the mixed state with
\eq{
\ket{\psi_{4,1}} = \ket{E_1}_{abc}\ket{0}_{d}\ket{0}_{e} + \ldots + \ket{E_1}_{bcd}\ket{0}_{a}\ket{3}_{e}
}
where the $\ldots$ entail all the other states associated with the remaining terms of the polynomial. Finally, the case $m=2$ is described by
\eq{\label{PN4k2}
\ma{P} = ab + ac + ad + bc + bd + cd\,,
}
whose associated link class can be represented by the link of Fig.~\ref{N4k2}.
The corresponding mixed state is readily obtained from
\eq{
\ket{\psi_{4,2}} = \ket{E_1}_{ab}\ket{00}_{d}\ket{0}_{e} + \ldots + \ket{E_1}_{cd}\ket{00}_{ab}\ket{5}_{e}\,.
}

\subsubsection{
 $N=5$}
\label{N5}

The case of $N=5$ can be analyzed in analogy to the previous cases.
The general $m$-resistant mixed state will be written as
\eq{\label{Gen5mixed}
\rrho(5,m) = {\textrm{tr}_{f} \left[\ket{\psi_{5,m}}\bra{\psi_{5,m}}\right] \over \sqrt{\braket{\psi_{5,m} | \psi_{5,m}}}}\,,
}
with $\ket{\psi_{5,m}}$ specified case by case. For $m=0$ we have
\eq{
\ma{P} = abcde
}
so the respective state will be
\eq{
\ket{\psi_{5,0}} = \ket{E_0}_{abcde}\ket{0}_{f}\,.
}
For $m=1$, we have
\eq{
\ma{P} = abcd + \ldots + bcde
}
which gives rise to the mixed state obtain from
\eq{
\ket{\psi_{5,1}} = \ket{E_1}_{abcd}\ket{0}_{e}\ket{0}_{f} + \ldots + \ket{E_1}_{bcde}\ket{0}_a\ket{4}_{f}\,.
}
For $m=2$, the polynomial is
\eq{
\ma{P} = abc + \ldots + cde\,,
}
which is mapped into the state
\eq{
\ket{\psi_{5,2}} = \ket{E_2}_{abc}\ket{00}_{de}\ket{0}_{f} + \ldots + \ket{E_2}_{cde}\ket{00}_{ab}\ket{9}_{f}\,.
}
Finally, for $m=3$, we have
\eq{
\ma{P} = ab + \ldots + de\,,
}
which gives the state
\eq{
\ket{\psi_{5,3}} = \ket{E_3}_{ab}\ket{000}_{cde}\ket{0}_{f} + \ldots + \ket{E_3}_{de}\ket{00}_{abc}\ket{9}_{f}\,.
}

\subsubsection{
 $N>5$}
\label{N6N7}

Taking into account the examples presented in the previous sections,
 it becomes tentative to conjecture the following statement.
 Let $S_{p-q}$ be the set containing all subsets of $p$ letters with $p-q$ elements, and let $\ma{B}_p$ be the set with all $p$ letters. If $N$ and $m$ are positive integers, with $N>m$, then the mixed state
\eq{\label{GenNrho}
\hat{\rho}(N,m) = {\textrm{tr}_{\Lambda} \left[\ket{\psi_{N,m}}\bra{\psi_{N,m}}\right] \over \sqrt{\braket{\psi_{N,m} | \psi_{N,m}}}}
}
with
\eq{\label{GenNket}
\ket{\psi_{N,m}} = \sum_{g_i \in S_{N-m}} \ket{E_m}_{g_i} \ket{0\ldots0}_{\ma{B}_N \backslash g_i} \ket{i}_{\Lambda}
}
is an $m$-resistant state of $N$ qubits. This result has been checked for $N$ up to $7$. For a larger number of qubits the size of the system grows as $2^N$
so the computing time required  to investigate separability of
 reduced density matrices increases accordingly.
  Therefore  studies of systems with a larger number of qubits,
  although possible,  becomes impracticable.

Regarding the representation of links with more than $6$ rings, it becomes less elucidating, so we will refrain from providing them in those cases. Nevertheless, using the respective polynomials, it is a straightforward task to draw these links.

\subsection{Pure qubit states}
\label{links}

Unlike the case with mixed states,  identification of a pure state of
the $N$-qubit system associated with a given $m$-resistant link
of $N$ rings is more intricate, if at all possible. We will provide here some partial answers following two different approaches: one which makes use of the polynomial
formalism for links; and another one based on the
Majorana representation~\cite{AMM10,MGBBB10}
of symmetric states of several qubits.

In the spirit similar to the previous sections we may inquire
if there exists a straightforward map between the polynomials of $m$-resistant links of $N$ rings,
and $m$-resistant pure states of $N$-qubits.
Taking into account Eqs.~(\ref{GenNrho}) and (\ref{GenNket}), the most immediate ansatz for a pure $m$-resistant $N$-qubit state, denoted as $\ket{(N,m)}$,
would be
\eq{\label{GenNpure}
\ket{(N,m)} = \sum_{g_i \in S_{N-m}} \ket{E_m}_{g_i} \ket{0\ldots0}_{\ma{B}_N \backslash g_i}\,.
}
Hence, we remove now the extra qudit in Eq.~(\ref{GenNket}) and
need not perform the partial trace. The state still possesses the same symmetries as the corresponding polynomial under particle permutations, so it is as least intuitive as a candidate. This ansatz,  tested up to $N=7$, gives a correct answer only for $m=0$, $N-1$ and $N-2$.
The first failure thus occurs for $N=5$ and $m=2$, where no corresponding $2$--resistant
pure state of $5$ qubits has yet been found. Other choices could be made for $\ket{E_m}$ in Eq.~(\ref{GenNpure}), for example, but as it turns out, no change in the coefficients of Eq.~(\ref{Em}) will provide the correct answer.

Different approaches can be used to search for $m$-resistant pure qubit states. Since the states can be symmetric under particle permutation, one may use the stellar
representation of Majorana \cite{AMM10,GKZ12},
which provides an alternative intuition on the geometry of these states.
The Majorana representation of a $N$-qubit pure state, fully symmetric under permutations of its $i_1, i_2,\ldots,i_N$ qubits, is given by
\eq{\label{MajRep}
\ket{\psi} = {1 \over \sqrt{K}} \sum_{\sigma \in \textrm{S}_N} \ket{\eta_1}_{i_1} \ldots \ket{\eta_N}_{i_N}
}
where the sum runs over all permutations $\sigma \in \textrm{S}_N$ of particles indices,
 $K$ is a suitable normalization constant and
\eq{
\ket{\eta_j}_{X} = \cos\left({\theta_j \over 2}\right)\ket{0}_X + e^{i \phi_j} \sin\left({\theta_j \over 2}\right) \ket{1}_X\,.
}
The $N$ pairs $(\theta_j,\phi_j)$ are called Majorana points and uniquely define a point in
 the Bloch sphere. In this way, one may define a fully symmetric $N$-qubit state by fixing $N$ points on the sphere. For this reason, the name stellar representation is also common,
 as each point represents a star in the sky,
 while a group of stars forms a constellation.

One of the advantages of the Majorana representation is that it allows one to ascribe
to entanglement a degree of geometrical intuition.
If one fixes all the stars at
a single point the corresponding state is separable.
For instance $N$ degenerated stars on the North pole
represent the  state $\ket{0\cdots0}$.
However, as the degeneracy is lifted, any non-trivial constellation
of stars corresponds to an entangled state of $N$ qubits.
Thus, in a sense, the distance between the stars is directly related to the degree of entanglement \cite{ZS01,GKZ12},
although such a criterium to quantify entanglement is  not uniquely defined.

\subsubsection{$N=3$}

\begin{figure}[h!]
  \subfloat[0-resistant state \label{N3m0Star}]{\includegraphics[width=.40\columnwidth]{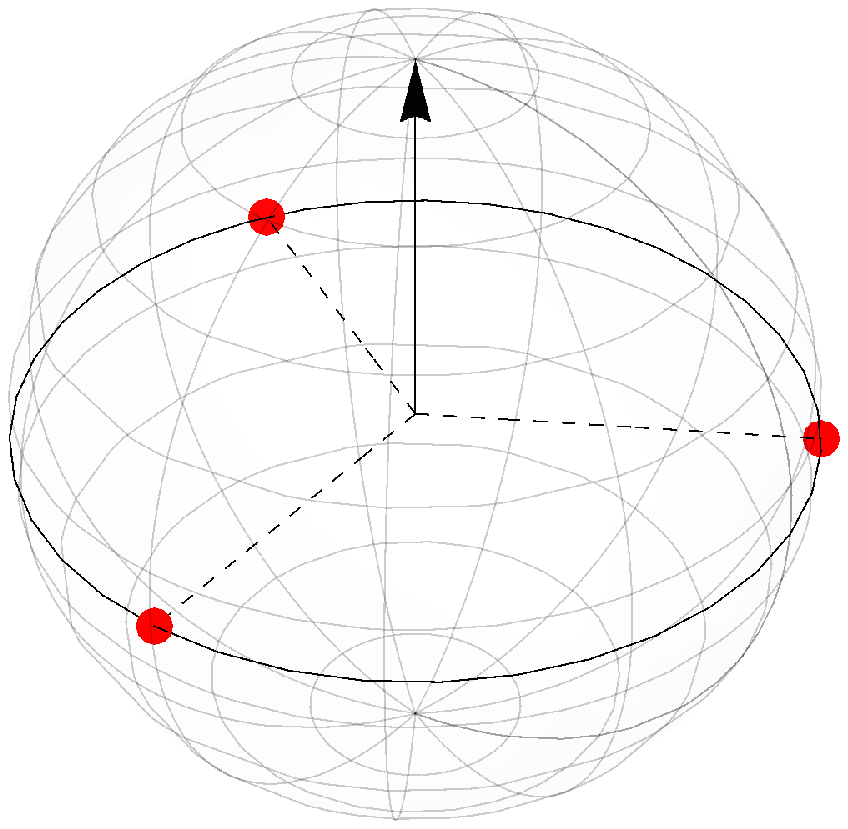}}
  \hspace{5mm}
  \subfloat[1-resistant state\label{N3m1Star}]{\includegraphics[width=.40\columnwidth]{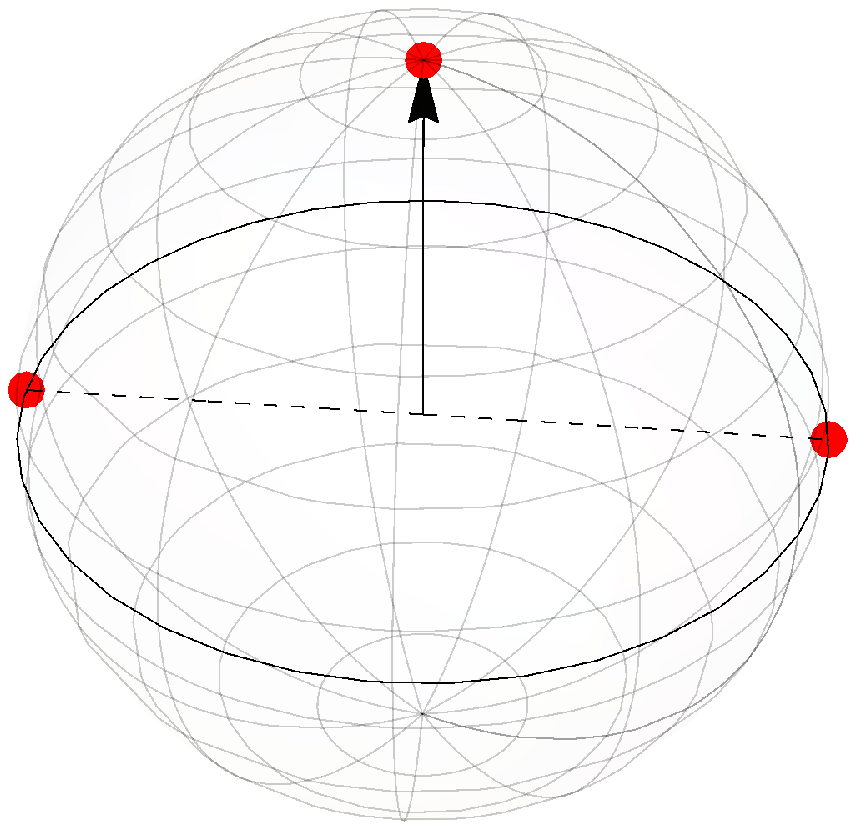}}
  \caption{Constellations defining $m$-resistant states of 3 qubits. The middle arrow serves as a reference pointing to the North pole. The constellation in a) corresponds to the state $|\textrm{GHZ}\rangle$ and is illustrated by the Borromean link in Fig. \ref{31}, and b) is related to the link in Fig. \ref{34}.}
  \label{StarsN3}
\end{figure}

In order to investigate possible patterns, one should start with the lowest qubit case $N=3$ and look for the simplest choices of constellations which give expected results. It is straightforward to show that the constellations of Fig.~\ref{StarsN3} give the desired answer for all $m$-resistant states of 3 qubits.
The $m$-resistant states of $N$-qubits constructed with help of
 the stellar representation
 and Majorana points will henceforth be denoted as $\ket{(N,m)_{\rm MP}}$. The constellation corresponding to $m=0$ results in the GHZ state
\eq{\label{N3m0State}
\ket{(3,0)_{\rm MP}} = {1\over \sqrt{2}}\left(\ket{000}+\ket{111}\right),
}
for which the distance between the points at the equator is the greatest.
From this perspective,  the $|\textrm{GHZ}\rangle$  state could be considered as the most entangled, although from the point of view of resistance,
entanglement of this $0$-resistant state is not resistant to the loss of a subsystem.
The case of $m=1$ corresponds to the state
\eq{
\ket{(3,1)_{\rm MP}} = {1\over \sqrt{12}}\left(3\ket{000}+\ket{011}+\ket{101}+\ket{110}\right)\,,
}
where the weight of the $\ket{000}$ is increased with respect
 to Eq.~(\ref{N3m0State}),  as one of the stars sits at the North pole.

\subsubsection{$N\geq 4$}

\begin{figure}[h!]
  \subfloat[$0$-resistant state\label{N4m0Star}]{\includegraphics[width=.3\columnwidth]{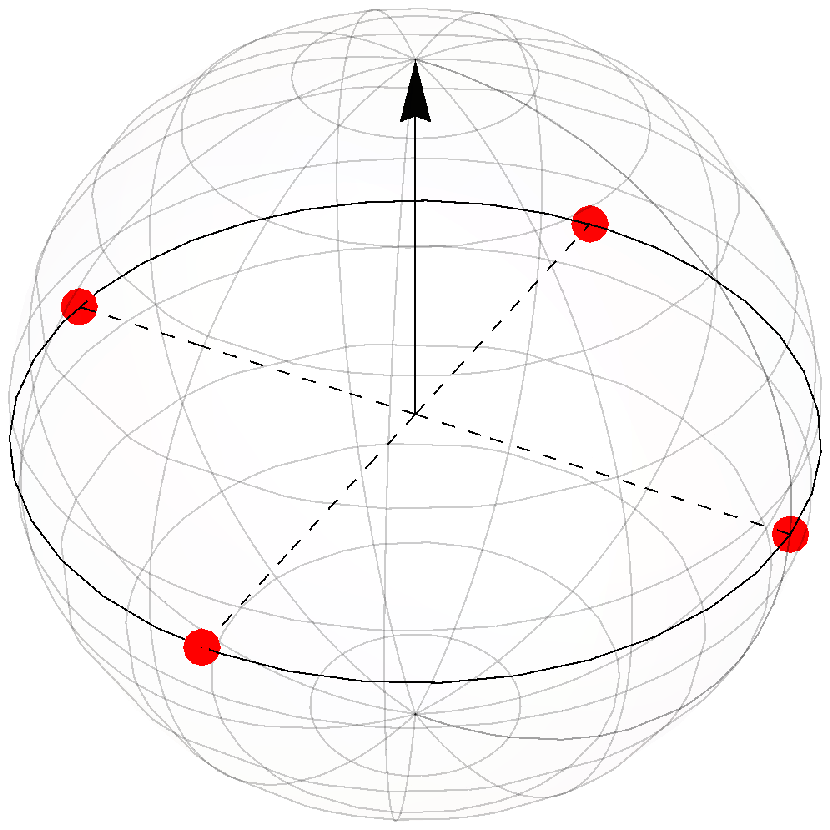}}
  \hfill
  \subfloat[$1$-resistant state\label{N4m1Star}]{\includegraphics[width=.3\columnwidth]{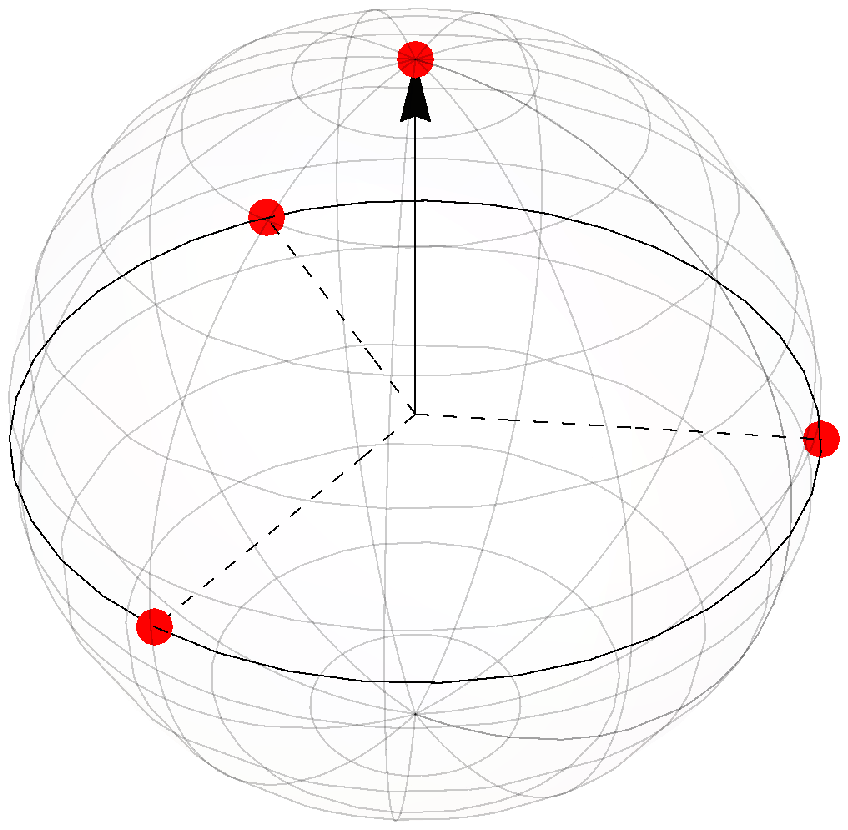}}
  \hfill
  \subfloat[$2$-resistant state\label{N4m2Star}]{\includegraphics[width=.3\columnwidth]{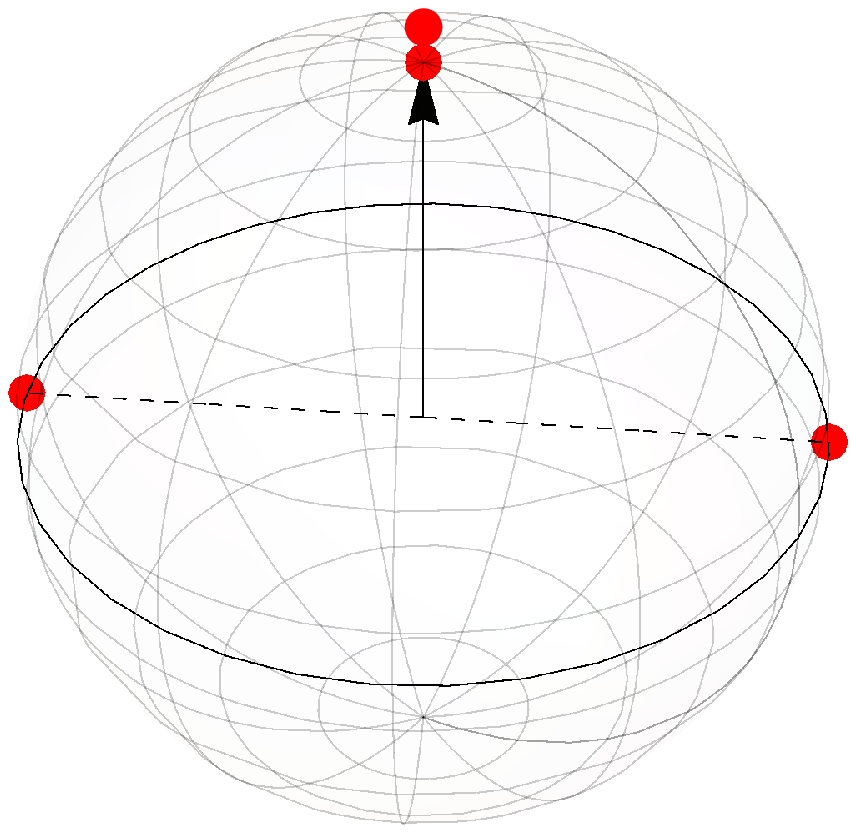}}
  \caption{Constellations defining $m$-resistant states of 4 qubits. The middle arrow serves as a reference pointing to the North pole. Constellation a) corresponds to the 4 ring Brunnian link from Fig. \ref{N4k0}, b) to the link in Fig. \ref{N4k1}, and c) to the one in Fig. \ref{N4k2}.}
  \label{StarsN4}
\end{figure}

The example of $N=3$ suggests that $m$-resistance may be directly related to the number of stars on the North pole, with all others evenly distributed along the equator. Indeed, for $N=4$ this is in fact the case, where pure qubit states for all $m$-resistant can be constructed exactly in this manner, with the constellations depicted in Fig.~\ref{StarsN4}.
The pattern is thus to place $m$ stars at the North pole and distribute the remaining stars evenly at the equator, where for convention we choose the first star at the equator with $({\pi \over 2},0)$. This construction gives rise to a family of states which, after expanding Eq.~(\ref{MajRep}), can be expressed in general as
\eq{\label{GenMPansatz}
\ket{\psi^{N}_{m}} = {1 \over \sqrt{1+\binom{N}{m}}}\left(\sqrt{\binom{N}{m}}\ket{0}^{\otimes N} - (-1)^{N+m} \ket{D^{N}_{m}}\right)
}
where we introduce the Dicke states \cite{Dicke}
\eq{
\ket{D^{N}_{m}} = {1 \over \sqrt{\binom{N}{m}}} \sum_{j} P_j \left\{ \ket{0}^{\otimes m} \ket{1}^{\otimes (N-m)} \right\}\,,
}
with the sum going through all permutations $P_j$. For $N=4$, we have
\ea{
\ket{\psi^{4}_{m}} = \ket{(4,m)}\,.
}

The construction described above breaks down at $N=5$.
 The first exception appears at $(5,1)$, where it is found that $\ket{\psi^{5}_{1}} = \ket{(5,2)}$. All other cases, i.e. $m=0$, 2 and 3, give pure states with
  the desired resistance. One may naturally inquire whether other symmetric positions of stars may give rise to a $\ket{(5,2)}$, other than 1 star on the North pole and 4 stars distributed in a square at the equator. We have tested combinations where two stars are lifted by a general latitude, as well as four stars, and it is possible to show that no $1$-resistant state exists for these families of states. Evidently, this does not prove that a pure $1$-resistant state of 5 qubits does not exist, although it is tantalizing to conjecture it, given that
pure states with certain entanglement properties under partial trace do not exist.
For instance, it is known \cite{HS00} that there are no AME states for four qubits.

For $N>5$ until $7$, we have verified that the ansatz in Eq.~(\ref{GenMPansatz}) fails in all situations except for $m=0$, $N-2$ and $N-1$. It is in fact possible to show for general $N$ that these cases are always satisfied, i.e. that
\ea{
\ket{\psi^{N}_{0}} & = \ket{(N,0)}\,, \\
\ket{\psi^{N}_{N-3}} & = \ket{(N,N-3)}\,, \\
\ket{\psi^{N}_{N-2}} & = \ket{(N,N-2)}\,.
}
A proof that $\ket{\psi^{N}_{0}}$ is zero resistant is straightforward. Indeed, by Eq.~(\ref{GenMPansatz}), we have
\eq{
\ket{\psi^{N}_{0}} = {1 \over 2}\left(\ket{0}^{\otimes N} - (-1)^{N} \ket{1}^{\otimes N}\right)\,.
}
This state is a generalization of the GHZ state for $N$ qubits, which, after any partial trace, returns a density matrix of the form
\eq{
{\textrm{tr}_{k}} \left[\hat{\rho}^{N}_{0}\right] = {1 \over 2}\left(\ket{0}\bra{0}^{\otimes (N-k)} + \ket{1}\bra{1}^{\otimes (N-k)}\right) ,
}
where ${\textrm{tr}_{k}}$ denotes the trace over any
set consisting of $k$ qubits.
This density matrix is separable for any $k>0$, and thus the state $\ket{\psi^{N}_{0}}$ is $0$-resistant for any $N$.

Regarding the state $\ket{\psi^{N}_{N-2}}$, after simple algebraic operations, one finds that any $N-2$ partial traces will result in the density matrix
\eq{
{\textrm{tr}_{N-2}} \left[ \hat{\rho}^{N}_{N-2} \right] =
\begin{pmatrix}
\alpha_{22} + (\alpha_{02})^2 & 0 & 0 & -\alpha_{02} \\
0 & \alpha_{21} & \alpha_{21} & 0 \\
0 & \alpha_{21} & \alpha_{21} & 0 \\
-\alpha_{02} & 0 & 0 & 1
\end{pmatrix} ,
}
where we define $\alpha_{ij} \equiv \binom{N-i}{j}$.
The partially transposed matrix has one of the eigenvalues equal to
\eq{
N^2+3N-4
}
which is negative for any $N$. Thus, by the
 positive partial transpose (PPT) test,
 we confirm that each 2-qubit reduced density matrix is always entangled, and the state $\ket{\psi^{N}_{N-2}}$ is $(N-2)$-resistant for any $N$.

Finally, focusing on the state $\ket{\psi^{N}_{N-3}}$, we
perform the partial trace over any set of  $N-2$ qubits and obtain
\eq{
{\textrm{tr}_{N-2}} \left[ \hat{\rho}^{N}_{N-3} \right] =
\begin{pmatrix}
\alpha_{23} + (\alpha_{03})^2 & 0 & 0 & 0 \\
0 & \alpha_{22} & \alpha_{22} & 0 \\
0 & \alpha_{22} & \alpha_{22} & 0 \\
0 & 0 & 0 & \alpha_{21}
\end{pmatrix} .
}
The partially transposed matrix of size four has all positive eigenvalues for $N>1$.
In this case the PPT test
 guarantees that the resulting state is always separable. We must then look into the reduced density matrix with one less partial trace, in order to check if it is entangled. Such density matrix has the form
\ea{
& {\textrm{tr}_{N-3}} \left[ \hat{\rho}^{N}_{N-3} \right] = \nonumber \\
& \begin{pmatrix}
\alpha_{33} + (\alpha_{03})^2 & 0 & 0 & 0 & 0 & 0 & 0 & \alpha_{03}\\
0 & \alpha_{32} & \alpha_{32} & 0 & \alpha_{32} & 0 & 0 & 0 \\
0 & \alpha_{32} & \alpha_{32} & 0 & \alpha_{32} & 0 & 0 & 0 \\
0 & 0 & 0 & \alpha_{31} & 0 & \alpha_{31} & \alpha_{31} & 0 \\
0 & \alpha_{32} & \alpha_{32} & 0 & \alpha_{32} & 0 & 0 & 0 \\
0 & 0 & 0 & \alpha_{31} & 0 & \alpha_{31} & \alpha_{31} & 0 \\
0 & 0 & 0 & \alpha_{31} & 0 & \alpha_{31} & \alpha_{31} & 0 \\
\alpha_{03} & 0 & 0 & 0 & 0 & 0 & 0 & 1 \\
\end{pmatrix}\,.
}
By partially transposing any qubit, we will obtain a matrix which possesses an eigenvalue equal to
\eq{
\left(13-7N+N^2 - \sqrt{193-202N + 79N^2 -14N^3 + N^4}\right)
}
which is negative for all $N$. PPT criterion implies that all the subsystems are entangled. This proves that the state $\ket{\psi^{N}_{N-3}}$ is $(N-3)$-resistant.

\section{In search for $m$-resistant qudit states}

As already  mentioned in \cref{links}, we were not able to establish
 a general construction of an $m$-resistant pure state of $N$ qubit system.
 Therefore, we have expanded our search for subsystems with a larger local dimension
 $d\ge 3$, sometimes called qudits. We succeeded in providing formulas
  for a $k$-resistant $N$-qudit pure state for $N \geq 2k$.

An example of the AME state of six qubits  \cite{BPBZCP07}
inspired us to search for states with diagonal density matrices after tracing the appropriate number of parts. More precisely, $m+1$ parts in order to find an $m$-resistant state. Trivially such states are separable after tracing $m+1$ parts.
Of course, the states we are looking for must remain entangled after tracing out a smaller number of parts.
In order to find such states, we present a connection between a family of combinatorial designs and a family of quantum states with required properties. In particular, we use the notion of \textit{orthogonal arrays} (\textit{OA}), and the established connection~\cite{Goyeneche:2014} between quantum states and OA.

We use the following, consistent with the hitherto, notation $\ket{(N,m)_d}$ for an $m$-resistant state of $N$ qudits.

\subsection{Orthogonal arrays and quantum states}
\label{reproduce}

Orthogonal arrays \cite{rao} are combinatorial arrangements, tables with entries satisfying given orthogonal properties. A close connection between OA and codes, entangled states, error-correcting codes, uniform states has been established \cite{OA}. Therefore,
investigation of the connections between OA and resistant states seems to be a
natural approach.
Firstly, the concept of OA is briefly presented; secondly, relations between OA and resistant states are given.

An orthogonal array $\oa{r,N,d,k}$ is a table composed by $r$ rows, $N$ columns with entries taken from $0,\ldots,d-1$ in such a way that each subset of $k$ columns contains all possible combination of symbols with the same amount of repetitions. The number of such repetitions is called \textit{the index} of the OA and denoted by $\lambda$. One may observe, that the index of OA is related to the previous coefficient by the following formula:
\eq{
\lambda =\dfrac{r}{d^k} .
}
Fig.~\ref{OA1} presents an example of an OA. A pure quantum state consisting of $r$ terms might be associated with $\oa{r,N,d,k}$, simply by reading all rows of OA
\cite{Goyeneche:2014}.
The state  of $N$ qudits
associated with the orthogonal array $\oa{r,N,d,k}$
will be denoted as
by $\ket{\phi_{(N,k )}}_d$.

\begin{figure}[h!]
\centering
\[
 \begin{array}{*5{c}}
    \tikzmark{left}{0} &
0&0&0&0 \\
0&1&1&1&1\\
0&2&2&2&2\\
0&3&3&3&3\\
1&0&1&2&3\\
1&1&0&3&2\\
1&2&3&0&1\\
1&3&2&1&0\\
2&0&2&3&1\\
2&1&3&2&0\\
2&2&0&1&3\\
2&3&1&0&2\\
3&0&3&1&2\\
3&1&2&0&3\\
3&2&1&3&0\\
3&    \tikzmark{right}{3}&0&2&1
\Highlight[first]
  \end{array}
  \qquad
  \begin{array}{*5{c}}
0&
\tikzmarkk{up}{0}&0&0&0 \\
0&1&1&1&1\\
0&2&2&2&2\\
0&3&3&3&3\\
1&0&1&2&3\\
1&1&0&3&2\\
1&2&3&0&1\\
1&3&2&1&0\\
2&0&2&3&1\\
2&1&3&2&0\\
2&2&0&1&3\\
2&3&1&0&2\\
3&0&3&1&2\\
3&1&2&0&3\\
3&2&1&3&0\\
3&  3&\tikzmarkk{down}{0}&2&1
\Highlightt[first]
  \end{array}
      \qquad
 \begin{array}{*3{c}}
 \ket{(5,1)_4}&=&\ket{00000} \\
 &+&\ket{01111} \\
&+&\ket{02222}\\
&+&\ket{03333}\\
&+&\ket{10123} \\
&+&\ket{11032}\\
&+&\ket{12301}\\
&+&\ket{13210}\\
&+&\ket{20231} \\
&+&\ket{21320} \\
&+&\ket{22013}\\
&+&\ket{23102}\\
&+&\ket{30312} \\
&+&\ket{31203} \\
&+&\ket{32130}\\
&+&\ket{33021}
  \end{array}
\]
\Highlightt[second]
\captionsetup{justification=justified,singlelinecheck=false}
\caption{Orthogonal array of unity index $\oa{4^2,5,4,2}$ obtained from Reed-Solomon code of length $5$ over GF($4$). Each subset consisting of two columns contains all possible combination of symbols. Here, two such subsets are highlighted. The relevant quantum state is obtained by forming a superposition
of states corresponding to consecutive rows of the array  -- see the expression
on the right-hand side.}
\label{OA1}
\centering
\end{figure}

\subsection{Orthogonal arrays of index unity}

The crucial quantity for our purpose, related to OA, is its index. It preserves the following information: how many repetitions of any sequence $i_1,\ldots,i_k$ there are for each subsystem of $k$ rows. For $\lambda =1$, any sequence appears only once, and such an array is called \textit{index unity array}. We emphasize their remarkable role in the search for resistant states.

\begin{proposition}
\label{prop1}
For any orthogonal array of index unity $\OA \left( d^k,N,d,k \right) $, where $N \geq 2k$, the relevant quantum state $\ket{\phi_{(N,k )}}$ is $k-1$-resistant. i.e
\eq{
\ket{\phi_{(N,k )}} =\ket{(N,k-1)_d}.
}
\end{proposition}

A proof of the latter statement is provided in Appendix \ref{appA}.
From the OA presented in Fig.~\ref{OA1}, we obtain, for example, the following
five-ququart, $1$-resistant state:
\begin{align}
\ket{(5,1)_4}=&\ket{00000} +\ket{01111} +\ket{02222}+\ket{03333}+ \nonumber \\
&\ket{10123} +\ket{11032} +\ket{12301}+\ket{13210}+ \nonumber \\
&\ket{20231} +\ket{21320} +\ket{22013}+\ket{23102}+ \nonumber \\
&\ket{30312} +\ket{31203} +\ket{32130}+\ket{33021}.
\end{align}

\subsection{Existence of OAs and relevant $m$-resistant states}

Construction of index unity OAs was provided by Bush \cite{Bush} in 1953. He used methods based on Galois fields theory in order to obtain the following result:

\begin{theorem}[Bush, 53']
If $d$ is a prime power, i.e $d=p^n$ for some prime number $p$ and natural number $n$, then we can construct the array $\OA \left( d^k, d+1,d,k\right) $.
\end{theorem}

Combining Bush's result with \cref{prop1} provides the existence of $m$-resistant $N$-qudit states for $N \geq 2(m+1)$.

\begin{corollary}
For any $N \geq 2(m+1)$ there exists the $N$-qudit state which is $m$-resistant. The local dimension $d$ is the smallest prime power larger than $N-1$.
\end{corollary}

In \cref{appB} we present some of the $m$-resistant qudit states. A more interested reader might easily reproduce more of them with the help of available OA tables \cite{tables}.

We organize all results obtained so far in \cref{table}. These results encourage us to pose the following conjecture:

\begin{conjecture}
For any $N$ and $m$, there exists an $m$-resistant $N$-qudit state in some local dimension $d$.
\end{conjecture}

\begin{table}
\begin{tikzpicture}[x=\daywidth, y=-1cm, node distance=0 cm,outer sep = 0pt]
\tikzstyle{day}=[draw, rectangle,  minimum height=1cm, minimum width=\daywidth, fill=yellow!15,anchor=south west]
\tikzstyle{day2}=[draw, rectangle,  minimum height=1cm, minimum width=1.5 cm, fill=yellow!20,anchor=south east]
\tikzstyle{hour}=[draw, rectangle, minimum height=1 cm, minimum width=1.5 cm, fill=yellow!30,anchor=north east]
\tikzstyle{1hour}=[draw, rectangle, minimum height=1 cm, minimum width=\daywidth, fill=yellow!30,anchor=north west]
\tikzstyle{Planche0}=[1hour,fill=red!40]
\tikzstyle{Planche1}=[1hour,fill=blue!50]
\tikzstyle{Planche2}=[1hour,fill=blue!35]
\tikzstyle{Planche3}=[1hour,fill=blue!20]
\tikzstyle{Planche4}=[1hour,fill=green!30]
\tikzstyle{Planche5}=[1hour,fill=blue!20]
\tikzstyle{Planche6}=[1hour,fill=blue!10]
\tikzstyle{Planche7}=[1hour,fill=blue!5]
\tikzstyle{Planche8}=[1hour,fill=blue!2]
\tikzstyle{Ang2h}=[2hours,fill=green!20]
\tikzstyle{Phys2h}=[2hours,fill=red!20]
\tikzstyle{Math2h}=[2hours,fill=blue!20]
\tikzstyle{TIPE2h}=[2hours,fill=blue!10]
\tikzstyle{TP2h}=[2hours, pattern=north east lines, pattern color=magenta]
\tikzstyle{G3h}=[3hours, pattern=north west lines, pattern color=magenta!60!white]
\tikzstyle{Planche}=[1hour,fill=white]
\node[day] (lundi) at (1,8) {4};
\node[day] (mardi) [right = of lundi] {5};
\node[day] (mercredi) [right = of mardi] {6};
\node[day] (jeudi) [right = of mercredi] {7};
\node[day] (vendredi) [right = of jeudi] {8};
\node[hour] (8-9) at (1,8) {0};
\node[hour] (9-10) [below = of 8-9] {1};
\node[hour] (10-11) [below= of 9-10] {2};
\node[hour] (11-12) [below = of 10-11] {3};
\node[hour] (12-13) [below  = of 11-12] {4};
\node[hour] (13-14) [below = of 12-13] {5};
\node[hour] (14-15) [below = of 13-14] {6};
\node[Planche1] at (1,8)  {$\ket{\psi^{4}_{0}}$}; 
\node[Planche2] at (1,9) {$\ket{\psi^{4}_{1}} $};
\node[Planche3] at (1,10) {$\ket{\psi^{4}_{2}} $};

\node[Planche1] at (2,8) {$\ket{\psi^{5}_{0}} $};
\node[Planche4] at (2,9) {$\ket{\phi^{}_{1,5}}_4 $};
\node[Planche2] at (2,10) {$\ket{\psi^{5}_{2}} $};
\node[Planche3] at (2,11) {$\ket{\psi^{5}_{3}} $};

\node[Planche1] at (3,8) {$\ket{\psi^{6}_{0}} $};
\node[Planche4] at (3,9) {$\ket{\phi^{}_{1,6}}_5 $};
\node[Planche0] at (3,10) {AME};
\node[Planche2] at (3,11) {$\ket{\psi^{6}_{3}} $};
\node[Planche3] at (3,12) {$\ket{\psi^{6}_{4}} $};

\node[Planche1] at (4,8) {$\ket{\psi^{7}_{0}} $};
\node[Planche4] at (4,9) {$\ket{\phi^{}_{1,7}}_7 $};
\node[Planche4] at (4,10) {$\ket{\phi^{}_{2,7}}_7 $};
\node[Planche2] at (4,12) {$\ket{\psi^{7}_{4}} $};
\node[Planche3] at (4,13) {$\ket{\psi^{7}_{5}} $};

\node[Planche1] at (5,8) {$\ket{\psi^{8}_{0}} $};
\node[Planche4] at (5,9) {$\ket{\phi^{}_{1,8}}_7 $};
\node[Planche4] at (5,10) {$\ket{\phi^{}_{2,8}}_7 $};
\node[Planche4] at (5,11) {$\ket{\phi^{}_{3,8}}_7 $};
\node[Planche2] at (5,13) {$\ket{\psi^{8}_{5}} $};
\node[Planche3] at (5,14) {$\ket{\psi^{8}_{6}} $};


\node[day2] at (1,8) {};
\draw (1,8)--(-0.25, 7);
\node at (0.65,7.5) {$N$};
\node at (0,7.5) {$m$};
\end{tikzpicture}
\captionsetup{justification=justified,singlelinecheck=false}
\caption{\label{table} Search for $m$-resistant states of  $N$ parties
at a glance.
Note that $\ket{\psi^{4}_{0}}$ is equivalent to
$|\textrm{GHZ}_4\rangle$ state of $4$ qubits.
Three families: $\ket{\psi^{N}_{0}} $, $\ket{\psi^{N}_{N-3}} $ and $\ket{\psi^{N}_{N-2}} $ of $m$-resistant $N$-qubit states discussed in \cref{search} are presented on the differently shaded blue background. Six-qubit state AME(6,2) is also signalized.
Moreover, $m$-resistant qudit states $\ket{\phi^{}_{m,N}}_d $ constructed by virtue of orthogonal arrays are demonstrated on the green background. The number in subscript is relevant to the local dimension $d$ of a state. Observe that the local dimension $d$ of qudit states is usually (but not always) equal to $N-1$.}
\end{table}


\section{Concluding remarks}
\label{conc}

Any  link of $N$ closed rings can be associated with a polynomial of $N$ variables.
This technique, developed  in \cite{QuintaAndre:2018},
allowed us to construct a class of $m$-resistant links of $N$ closed rings
such that, after cutting any number $m$ of them, the remaining rings are connected,
while cutting any of the remaining rings separates them.
Making use of the analogy between
linked rings and  entangled states \cite{AS07}
we provided examples of  $m$-resistant entangled
quantum states, such that their entanglement is resistant to the loss of
any set of $m$ subsystems.

An explicit construction involving the partial trace
provides examples of $m$-resistant mixed states of $N$-qubit systems.
An alternative method based on stellar representation yields
$m$-resistant pure states, but for $N\ge 5$ it does
not produce all the required states for all values of $m=0,1,\dots N-2$.
However, in some cases for which the above method fails
(e.g. $N=5$ and $m=1$) we constructed $m$-resistant pure
states of $N$ parties with a higher local dimension $d$
making use of the combinatorial notion of orthogonal arrays~\cite{OA}.

It is important to emphasize two defining characteristics of any $m$-resistant quantum state.
The first one is the presence or disappearance of entanglement after partial traces, which is equivalent to fixing which subsystems of a quantum state are entangled, when the remaining part of the system is ignored or cannot be measured. In other words, it fixes which subsystems are able to successfully perform protocols between each other. The second defining property is the symmetry under permutations of parties, which brings about a notion of equality between all subsystems.

These two characteristics of quantum resistance may be combined into applications which require the simultaneous action of any given number of parties in order to perform some concrete task. For instance, consider a digital locker belonging to a certain network of $N$ parties, where each person is in possession of a qubit,
 which is part of a genuinely entangled $N$ qubit state.
 One may devise this locker such that it may only be opened through a protocol involving at least $m$ people. In this case, an $m$-resistant state of $N$ qubits will provide exactly this conditions, since any $m-1$ parties will always be in a separable state that makes it impossible to form correlations between them.

Finally, we leave as a final remark a list of relevant open problems:
a) check whether there exist pure states of $N$ qubits
with the $m$-resistance property, for any $m=0,1,\dots N-2$;
b) if this is not the case, for each $N$ find the minimal local dimension $d$
such that there exist $m$-resistant states of $N$ qudits; and
c) for any class of $m$-resistant states of $N$ qubits find a state for which
its average entanglement after partial trace over any set of $m$ parties
is the largest,  if measured with respect to a given measure of entanglement.

\acknowledgements
The authors would like to thank Miguel Reis Orcinha for helping
with computational aspects of the code used for testing the separability of the density matrices.
It is a pleasure to thank Ole Andersson, Ingemar Bengtsson and
 Pawe{\l} Horodecki for numerous fruitful dicussions and to
Konrad Szyma{\'n}ski for designing Fig.~12.
G.Q. acknowledges support from Funda\c{c}\~{a}o para a Ci\^{e}ncia e a Tecnologia (Portugal) through programme POCH and project UID/EEA/50008/2019.
R.A. acknowledges support from the Doctoral Programme in the Physics and Mathematics of Information (DP-PMI) and FCT through Scholarship No.~PD/BD/135011/2017.
A.B. and K.\.Z.
are supported by National Science Center under the
grant number DEC-2015/18/A/ST2/00274.

\appendix

\section{Proof of \cref{prop1}}
\label{appA}

\begin{proof}[Proof of \cref{prop1}]
In order to prove the $(k-1)$-resistance of $\ket{\phi_{(N,k )}} $ we need to show that it becomes separable after the partial trace of any $k$ particles; while it remains entangled as any $k-1$ particles are traced away.

Observe that after tracing out any $k$ particles, the density matrix becomes a diagonal matrix, hence it represents a separable state.

Checking the entanglement properties of the system after tracing $k-1$ particles is a bit more tricky.
We shall show that such a system after partial transpose of one particle will always have a negative eigenvalue.
Hence by PPT criterion,  it remains entangled.

We investigate the density matrix $\rho_a (\psi)$ after tracing out first $(k-1)$ particles.
We will show that the partial transpose of the last particle $\rho_{k-1}^{T_N} (\psi)$ have non-negative eigenvalues.
To clarify, we sketch an example in Fig.~{OA} of how the argument for the proof is developed.

\begin{figure}[h]
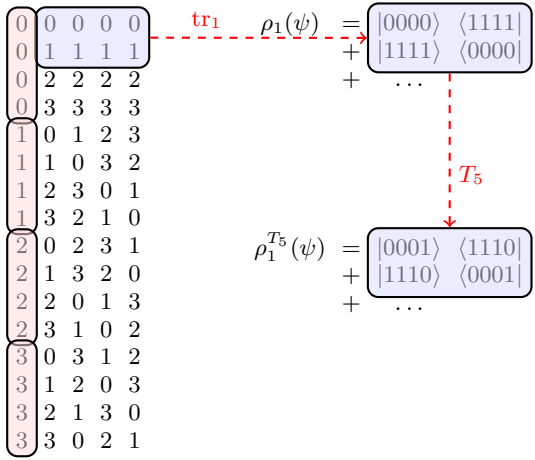

\[
 \begin{array}{*5{c}}
    \tikzmark{left}{0} &
\tikzmarkk{up}{0}&0&0&0 \\
0&1&1&1&\tikzmarkk{down}{1}\\
\Highlightt[firstt]
0&2&2&2&2\\
\tikzmark{right}{0}&3&3&3&3\\
 \Highlight[second]
 \tikzmark{left}{1}&0&1&2&3\\
1&1&0&3&2\\
1&2&3&0&1\\
\tikzmark{right}{1}&3&2&1&0\\
 \Highlight[first]
 \tikzmark{left}{2}&0&2&3&1\\
2&1&3&2&0\\
2&2&0&1&3\\
\tikzmark{right}{2}&3&1&0&2\\
 \Highlight[first]
 \tikzmark{left}{3}&0&3&1&2\\
3&1&2&0&3\\
3&2&1&3&0\\
\tikzmark{right}{3}&3&0&2&1
 \Highlight[first]
  \end{array}
  \qquad
      \qquad
 \begin{array}{*4{c}}
 \rho_{1} (\psi) &=   &\tikzmark{up}{$\ket{0000}$}& \bra{111 1} \\
   &+& \ket{1 1 1 1} &\tikzmark{down}{$\bra{000 0}$} \\
   &+& \ldots &\\
\Highlightt[first]
   \\
   \\
   \\
   \\
   \\
 \rho_{1}^{T_5} (\psi) &=   &\tikzmarkk{up}{$\ket{0001}$}& \bra{111 0} \\
   &+& \ket{1 1 1 0} &\tikzmarkk{down}{$\bra{000 1}$} \\
   &+& \ldots \\
   \\
   \\
   \\
   \\
   \\
  \end{array}
\]
\Highlightt[second]

\tikz[overlay,remember picture] {
  \draw[->,thick,red,dashed] (first) -- (second) node [pos=0.66,right] {$ T_5$};
  \node[above of=first] {};
  \node[above of=second] {};
}
\tikz[overlay,remember picture] {
  \draw[->,thick,red,dashed] (firstt) -- (first) node [pos=0.26,above] {$ \tr_1$};
  \node[above of=first] {};
  \node[above of=second] {};
}
\captionsetup{justification=justified}
\caption{\label{OA} Orthogonal array of unity index $\oa{4^2,5,4,2}$.
We investigate the matrix $\rho_{1}^{T_5} (\psi)$ obtained by tracing out the first particle and transposing the last one.
The first two rows of presented OA differ in all positions except for the first one.
They generate the elements $ \ket{0001} \bra{111 0}$ and $\ket{1 1 1 0} \bra{000 1}$ in $\rho_{1}^{T_5} (\psi)$.
There are no other elements in $\rho_{1}^{T_5} (\psi)$ involving $\ket{0001} ,\ket{1 1 1 0}$, and $\bra{111 0}, \bra{000 1}$.
Consequently, the matrix $\rho_{1}^{T_5} (\psi)$ has a negative eigenvalue. A similar argument holds for any other choice of a traced particle, and thus $\rho_1 (\psi)$ is an entangled state.}
\end{figure}

To begin, take two rows of the OA, which have zeros on the first $(k-1)$ positions.
Since the OA has strength $k$ and is of index unity, those rows differ in all other positions.
Without loss of generality assume that they are of the form
\[
 \begin{array}{*4{c}}
&0\cdots 0&0\cdots 0 &\\
&\underbrace{0\cdots 0}_{k-1}& \underbrace{1\cdots 1}_{N-k+1} &.\\
 \end{array}
\]
Consider now the density matrix $\rho_{k-1} (\psi)$ obtained by tracing out first $(k-1)$ particles, and examine the eigenvalues of the partial transpose of the last particle $\rho_{k-1}^{T_l} (\psi)$.
The aforementioned rows generate the following elements in the matrix $\rho_{k-1}^{T_l} (\psi)$:
\[
\begin{matrix}
\ket{0 \cdots 0 1} \bra{1\cdots 1 0},\\
 \ket{1 \cdots 1 0} \bra{0\cdots0 1} .
\end{matrix}
\]

We shall show that those elements are the only ones interlocking the bras
\[
\bra{1\cdots 1 0}, \bra{0\cdots0 1}
\]
and kets
\[
\ket{0 \cdots 0 1} ,\ket{1 \cdots 1 0} .
\]
With this observation at hand, one can see that the matrix $\rho_{k-1}^{T_l} (\psi)$ has the block structure:
\[
\rho_{k-1}^{T_l} (\psi)=
\begin{pmatrix}
&   \tikzmark{left}{0} &1&0&\cdots&0 &\\
&1&   \tikzmark{right}{0} &0&\cdots&0 &\\
\Highlight[first]
&0&0&   \tikzmark{left}{$\cdot$} &\cdots&\cdot& \\
&\vdots  & \vdots& \vdots &\vdots \vdots \vdots &\vdots &\\
&0&0&\cdot&\cdots&   \tikzmark{right}{$\cdot$}& \\
\Highlight[first]
\end{pmatrix}
,
\]
where we changed the order of computational basis in such a way that the first two vectors are
\[
 \begin{array}{*3{c}}
&\ket{0\cdots 0}&\\
&\underbrace{\ket{1\cdots 1}}_{N-k+1} &.
 \end{array}
\]
Consequently, $\rho_{k-1}^{T_l} (\psi)$ has at least one negative eigenvalue, namely $-1$.

It only remains to show the block form of the matrix $\rho_{k-1}^{T_l} (\psi)$. One may deduce it from the following two observations:
\begin{enumerate}
\item The first row in OA is the only one having zeros at positions $k,\ldots , N$ among the first $d$ rows. Indeed, each of the first $d$ rows has zeros at the first $k-1$ position. Since the OA is of unity index at each other position, the elements will differ between rows;
\item  There are no other rows having only zeros or only ones at the positions $k,\ldots N-1$. Indeed, since we assumed $N \geq 2k$, the number of those positions is $\geq k$. The first and the second rows have zeros, and ones, respectively, in those positions. Since the OA is of unity index, no other row may inherit this property.
\end{enumerate}
This ends the proof.
\end{proof}

\section{Examples of resistant states obtained from OA}
\label{appB}

We present some of the $m$-resistant qudit states obtained using the established connection with orthogonal arrays. More states can be straightforwardly found with the help of available OA tables in \cite{tables}.

The state $\ket{(7,1)_7}$ might be obtained form $\ket{(8,1)_7}$ simply by deleting the last qudit from each summand.
The number of summands grows rapidly as a function $N^{m=1}$.
For instance, the states $\ket{(7,2)_7}$ and $\ket{(8,2)_7}$ are superpositions of $7^3$ elements each; while the state $\ket{(8,3)_7}$ consists of $7^4$ elements.
The states can be quickly obtained through the method demonstrated in \cref{reproduce} with the help of available OA tables in \cite{tables}.

\begin{align}
\ket{(6,1)_5}=&\ket{000000} +\ket{011234} +\ket{022341}+\ket{033412}+ \nonumber \\
&\ket{044123} +\ket{101111} +\ket{112403}+\ket{124032}+ \nonumber \\
&\ket{130324} +\ket{143240} +\ket{202222}+\ket{214310}+ \nonumber \\
&\ket{223104} +\ket{231043} +\ket{240431}+\ket{303333} + \nonumber \\
&\ket{310142} +\ket{321420} +\ket{334201}+\ket{342014}+ \nonumber \\
&\ket{404444} +\ket{413021} +\ket{420213}+\ket{432130} + \nonumber \\
& \ket{441302}.
\end{align}

\begin{widetext}
\begin{align}
\ket{(8,1)_7}=&\ket{00000000} +\ket{01123456} +\ket{02234561}+\ket{03345612}+\ket{04456123} +\ket{05561234} +\ket{06612345}+ \nonumber \\
&\ket{10111111}+ \ket{11352064} +\ket{12520643} +\ket{13206435}+\ket{14064352}+ \ket{15643520} +\ket{16435206} + \nonumber \\
&\ket{20222222} +\ket{21546301} +\ket{22463015}+\ket{23630154}+\ket{24301546} +\ket{25015463} +\ket{26154630}+\nonumber \\
&\ket{30333333}+ \ket{31265140} +\ket{32651402} +\ket{33514026}+\ket{34140265}+ \ket{35402651} +\ket{36026514} + \nonumber \\
&\ket{40444444}+ \ket{41031625} +\ket{42316250} +\ket{43162503}+\ket{44625031}+ \ket{45250316} +\ket{46503162} + \nonumber \\
&\ket{50555555}+ \ket{51604213} +\ket{52042136} +\ket{53421360}+\ket{54213604}+ \ket{55136042} +\ket{56360421} + \nonumber \\
&\ket{60666666}+ \ket{61410532} +\ket{62105324} +\ket{63053241}+\ket{64532410}+ \ket{65324105} +\ket{66241053} .
\end{align}
\end{widetext}



\section{$m$-resistant links revisited}

The notion of $m$-resistant links is shown to be useful
to visualize entanglement properties of a multipartite quantum state.
Furthermore, it can be also relevant in the context of knot theory.
Even though from this perspective
it is not  common to analyze problems related to cutting the links or neglecting them,
the literature contains some items on  this topic \cite{Ohyama}.

We emphasize the relation between $m$-resistance and $n$-triviality of a link diagram.
Let us begin with the precise definition of an $m$-resistant link.
\begin{definition}[$m$-resistant link]

\label{mres}

A connected link consisting  of $N$ components is called $m$-resistant if:

\begin{itemize}

\item it remains linked as any $m$ components are neglected;

\item it becomes unlinked after neglecting any $m + 1$ components.

\end{itemize}

\end{definition}
The most common example of $m$-resistant link is the \textit{Brunnian link}, with $m=0$. We recall that the Brunnian link is a non-trivial link that becomes a set of trivial unlinked rings if any single component is removed.

A Brunnian link consisting of $N$ rings is by construction
$0$-resistant,
 see Fig.~\ref{N4k0}. There is a natural question, which is whether, for any $N$ and $m<N$, there exists an $m$-resistant link or not. We shall show that is in fact possible, as it should be on intuitive grounds.

\begin{proposition}

For any $N$ and $m= 0,1\dots, N-1$, there exists an $m$-resistant link of $N$ rings.

\end{proposition}

To prove this statement we use the technique of associating rings to polynomials
provided in \cite{QuintaAndre:2018}.
  Consider a polynomial of $N$ variables, $P_m(x_1,x_2,\dots,x_N)$,
  with  $\binom{N}{m}$ different terms,
  each being a product of $N-m$ variables. If $m+1$ variables are set
  to zero, then $P_m$ is equal to zero, which is taken to imply
  that the corresponding link becomes disjoint after cutting $m+1$ links.

\begin{figure}[h!]
  \centering

  \vspace{2mm}
  \subfloat[Braid representation for 3 ring Brunnian link.\label{3braid}]{\includegraphics[width=0.45\textwidth]{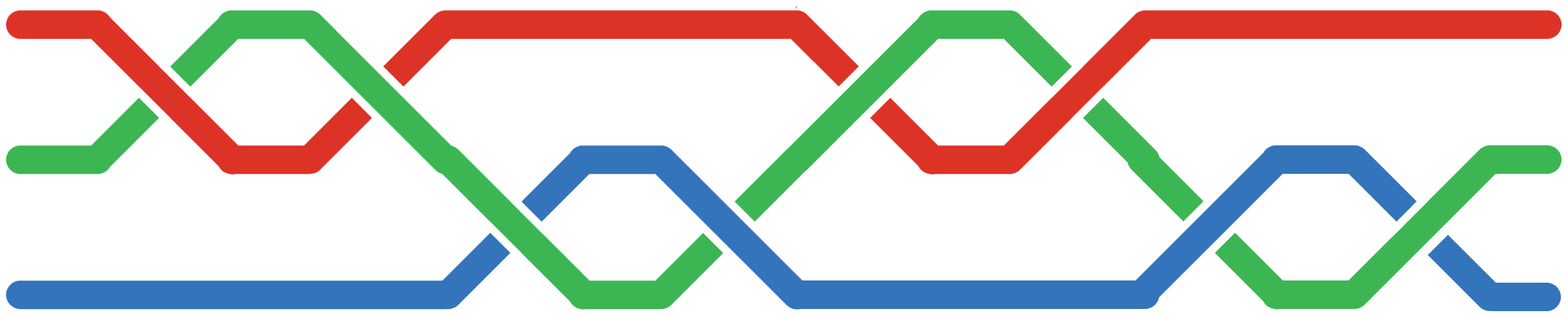}}\\
  \vspace{2mm}
  \subfloat[Braid representation for 4 ring Brunnian link.\label{4braid}]{\includegraphics[width=0.9\textwidth]{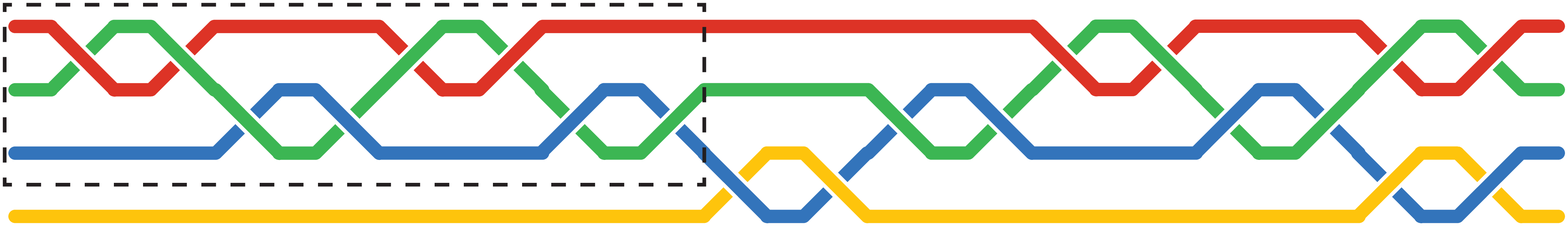}}
      \caption{The braid representations for Brunnian links of 3 rings (top) and 4 rings (bottom). These braids constitute the building blocks for $(N-4)$ and $(N-3)$-resistant links on $N$ rings respectively. In the braid representation, the links are obtained by joining each end of the braids. It is interesting to note that, inside the dashed box in b), one has the braid representation of the 3 ring Brunnian link of a).}
      \label{braids}

\end{figure}

The procedure for constructing such links uses Brunnian braids spanned on the appropriate number of rings as basic building blocks. In particular, constructing an $m$-resistant link of $N$ components amounts to constructing a polynomial of $N$ variables with $\binom{N}{N-m}$ terms of $N-m$ variables each, and identifying each of these terms with the appropriate Brunnian building block -- see Fig.~\ref{braids}. Each Brunnian building block is a pure braid, i.e. a braid where the lines begin and end in the same order. In addition, each Brunnian braid interlaces only the lines corresponding to the variables appearing in the specific term of the polynomial, with the remaining lines present but not interlaced. In the end, all braids are put together by connecting lines of the same color to form a single braid, corresponding to the specific $m$-resistant link obeying both conditions required in \cref{mres}. The order by which they are connected does not matter.

\begin{figure}[h!]
  \centering
  \includegraphics[width=0.70\textwidth]{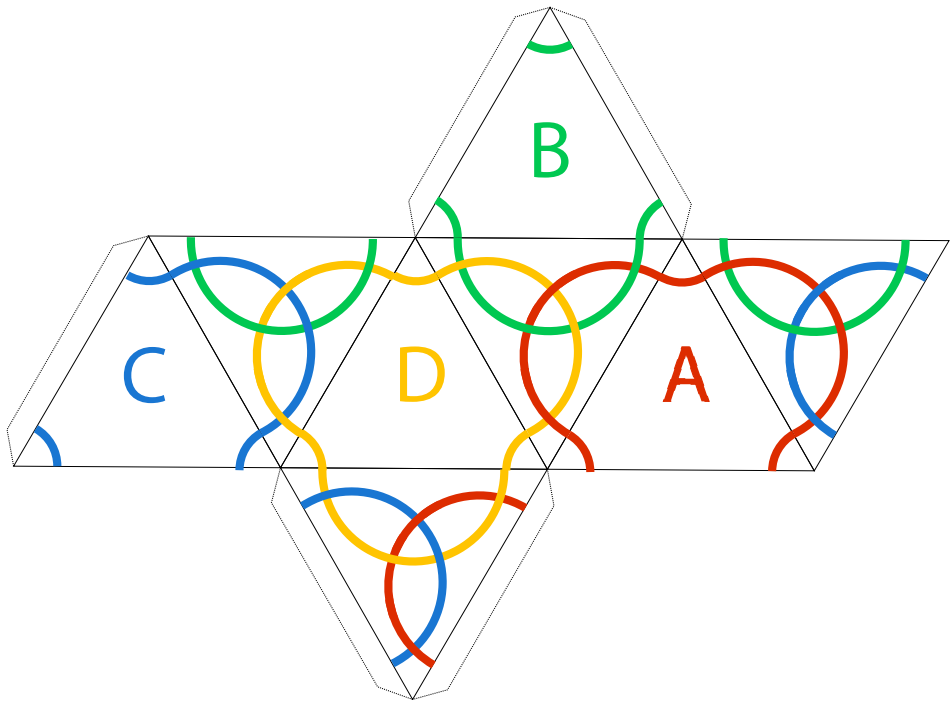}
      \caption{The blue print of an octahedron corresponding to the 1-resistant link of 4 rings (see Fig.~\ref{N4k1}), classified by the polynomial ${\red a}{\green b}{\blue c} + {\red a}{\green b}{\yellow d} + {\red a}{\blue c}{\yellow d} + {\green b}{\blue c}{\yellow d}$, given in Eq.~(\ref{PN4k1}). There is one face for each term of the polynomial with a Brunnian link of 3 rings with open edges of all rings. When the edges of the blue print are glued together, the edges of each ring are connected according to their color, without any extra crossings.}
      \label{simplex}

\end{figure}

Finally, the procedure which is used to represent all $m$-resistant links in this work, developed in \cite{QuintaAndre:2018}, can also be adapted in a way which is easier for 3D visualization. For a given $m$-resistant link, this procedure begins by drawing one Brunnian link for each term of the polynomial that characterizes the link and then cutting one end of each ring. The ends of all rings of the same color are then connected in such a way that no additional crossing are performed. A particular way to systematize this process is by building blueprints like the one described in Fig.~\ref{simplex}.

\onecolumngrid

\twocolumngrid

\end{document}